\documentclass{article} %
\usepackage{iclr2025_conference,times}

\usepackage{adjustbox} %
\usepackage{amsfonts}
\usepackage{amsmath}

\usepackage{amssymb}
\usepackage{array} %
\usepackage{bbding}
\usepackage{booktabs} %
\usepackage{colortbl}
\usepackage{graphicx} %
\usepackage{hyperref}
\usepackage{makecell}
\usepackage{multicol} %
\usepackage{multirow}
\usepackage{setspace}
\usepackage{tcolorbox} %
\usepackage{url}

\usepackage{ulem}
\usepackage{pifont}
\usepackage{wrapfig}
\usepackage{xcolor}
\usepackage{xspace}
\usepackage{mathrsfs} %
\usepackage{graphicx}
\usepackage{subcaption}
\usepackage{float}
\usepackage{CJKutf8}

\usepackage{hyperref}
\hypersetup{
  colorlinks   = true,    %
  urlcolor     = blue,    %
  linkcolor    = blue,    %
  citecolor    = blue      %
}

\newcolumntype{C}[1]{>{\centering\arraybackslash}p{#1}}

\newcommand{\eos}{\texttt{<EOS>}\xspace}

\newcommand{\unk}{\texttt{<UNK>}\xspace}
\definecolor{framec}{RGB}{62,62,62}

\definecolor{backc}{RGB}{242,242,242}

\newenvironment{packeditemize}{
\begin{list}{$\bullet$}{
\setlength{\labelwidth}{8pt}
\setlength{\itemsep}{0pt}
\setlength{\leftmargin}{\labelwidth}
\addtolength{\leftmargin}{\labelsep}
\setlength{\parindent}{0pt}
\setlength{\listparindent}{\parindent}
\setlength{\parsep}{0pt}
\setlength{\topsep}{3pt}}}{\end{list}}

\usepackage{amsmath,amsfonts,bm}

\def\eqref#1{equation~\ref{#1}}

\def\1{\bm{1}}

\DeclareMathAlphabet{\mathsfit}{\encodingdefault}{\sfdefault}{m}{sl}
\SetMathAlphabet{\mathsfit}{bold}{\encodingdefault}{\sfdefault}{bx}{n}

\title{An Engorgio Prompt Makes Large Language Model Babble on}

\author{
    \centerline{Jianshuo Dong\textsuperscript{1}, \ 
    Ziyuan Zhang\textsuperscript{1}, \ 
    Qingjie Zhang\textsuperscript{1}, \ 
    Tianwei Zhang\textsuperscript{2},} \vspace{0.5mm}  \\
    \centerline{\textbf{\ Hao Wang\textsuperscript{1}, \
    Hewu Li\textsuperscript{1}, \ 
    Qi Li\textsuperscript{1}, \ 
    Chao Zhang\textsuperscript{1}, \ 
     Ke Xu\textsuperscript{1},\ 
    and Han Qiu\textsuperscript{1}\thanks{The corresponding author}
    }} \vspace{0.5mm} \\
    \centerline{\normalsize{$^{1}$Tsinghua University, \ \ \  $^{2}$Nanyang Technological University}} \vspace{0.5mm} \\
    \centerline{\texttt{\small dongjs23@mails.tsinghua.edu.cn}, \ \ \
    \texttt{\small qiuhan@tsinghua.edu.cn}}
}

\iclrfinalcopy % Uncomment for camera-ready version, but NOT for submission.

\begin{document}

\maketitle

\vspace{-1em}
\begin{abstract}
Auto-regressive large language models (LLMs) have yielded impressive performance in many real-world tasks. 
However, the new paradigm of these LLMs also exposes novel threats. 
In this paper, we explore their vulnerability to inference cost attacks, where a malicious user crafts Engorgio prompts to intentionally increase the computation cost and latency of the inference process. We design Engorgio, a novel methodology, to efficiently generate adversarial Engorgio prompts to affect the target LLM's service availability. Engorgio has the following two technical contributions. 
(1) We employ a parameterized distribution to track LLMs' prediction trajectory. (2) Targeting the auto-regressive nature of LLMs' inference process, we propose novel loss functions to stably suppress the appearance of the \eos token, whose occurrence will interrupt the LLM's generation process. 
We conduct extensive experiments on 13 open-sourced LLMs with parameters ranging from 125M to 30B. 
The results show that Engorgio prompts can successfully induce LLMs to generate abnormally long outputs (i.e., roughly 2-13$\times$ longer to reach 90\%+ of the output length limit)
in a white-box scenario and our real-world experiment demonstrates Engergio's threat to LLM service with limited computing resources.
The code is released at: \url{https://github.com/jianshuod/Engorgio-prompt}. 
\end{abstract}

\section{Introduction}\label{sec:intro}

Large language models (LLMs)~\citep{touvron2023llama,ouyang2022training,carlini2021extracting} have demonstrated remarkable performance in various real-world applications, e.g., online chatting~\citep{shen2023chatgpt}, customer service~\citep{gimpel2023unlocking}, and finance~\citep{wu2023bloomberggpt}.
Given the increasing popularity and adoption of LLMs, reducing their inference cost becomes critical. 
Firstly, from the \textit{cost} aspect, a modern LLM normally contains billions of parameters, and each inference generation may consume considerable resources and time. 
Many AI service providers are paying more bills to support their LLM inference services than training~\citep{patel2024splitwise,li2024llm, patterson2022carbon}. 
Secondly, from the \textit{service availability} aspect, there is fierce competition across different LLM service providers, making service reliability and fast response time important factors in attracting customers. 
Meanwhile, these two considerations motivate malicious entities to attack the LLMs, increasing their operational cost and generation latency. 

In this paper, we explore the landscape of \textbf{inference cost attacks} against modern LLMs. 
First proposed in \citet{shumailov2021sponge} to attack encoder-decoder transformers, inference cost attacks aim to intentionally maximize the energy consumption and latency of model inference via a new type of adversarial input. 
The inference cost attacks on language models \citep{shumailov2021sponge,chen2022nmtsloth,feng2024llmeffichecker} are tailored for encoder-decoder models and rely on perturbation-based mutation to progressively hit a desirable adversarial input.
However, as demonstrated in Section~\ref{sec:relatedWorks} and Section~\ref{sec:exp}, they become ineffective against modern LLMs~\citep{autoregressive}, which adopt the auto-regressive generation scheme~\citep{autoregressive}, remove the cross-attention mechanism, and employ a sub-word tokenization algorithm. 
\citet{geiping2024coercing} propose an adversarial prompt attack to coerce LLMs into repeating specific content,  achieving effects similar to an inference cost attack. However, its reliance on the starting response weakens robustness and increases detectability.

In general, it is challenging to design inference cost attacks against modern decoder-only LLMs, even given the existing works discussed above.
The main challenges stem from two aspects: \textbf{(1)} \textbf{Uncertain Generation Process.}
The generation process of decoder-only LLMs is inherently casual, auto-regressive, and sampling-based, rendering it difficult to constrain them to generate a specific long response.
The occurrence of one deviant token can directly distort the generation process from the desirable decoding trajectory, challenging attack effectiveness and stability.
\textbf{(2)} \textbf{Discrete Input Modality.}
Text-completion LLMs accept input text in the form of discrete token sequences but operate within the embedding space, which implies an irreversible mapping from the embedding space back to the token space.
While we can leverage gradient information to optimize a more desirable soft embedding representation for the input, we face challenges in accurately identifying corresponding tokens in the token space for the optimized soft embeddings.
This restricts us from effectively leveraging gradients to guide updates to the input token sequence (i.e., adversarial input).
To address the above challenges, we need to consider two intriguing and important questions: \textbf{(1)} how to accurately frame our goal as a well-aligned optimization problem and \textbf{(2)} how to effectively instruct the updates to the discrete input sequence given the modeled objective.

\begin{wrapfigure}{r}{0.53\linewidth}
    \centering
\vspace{-1em}
\includegraphics[width=\linewidth]{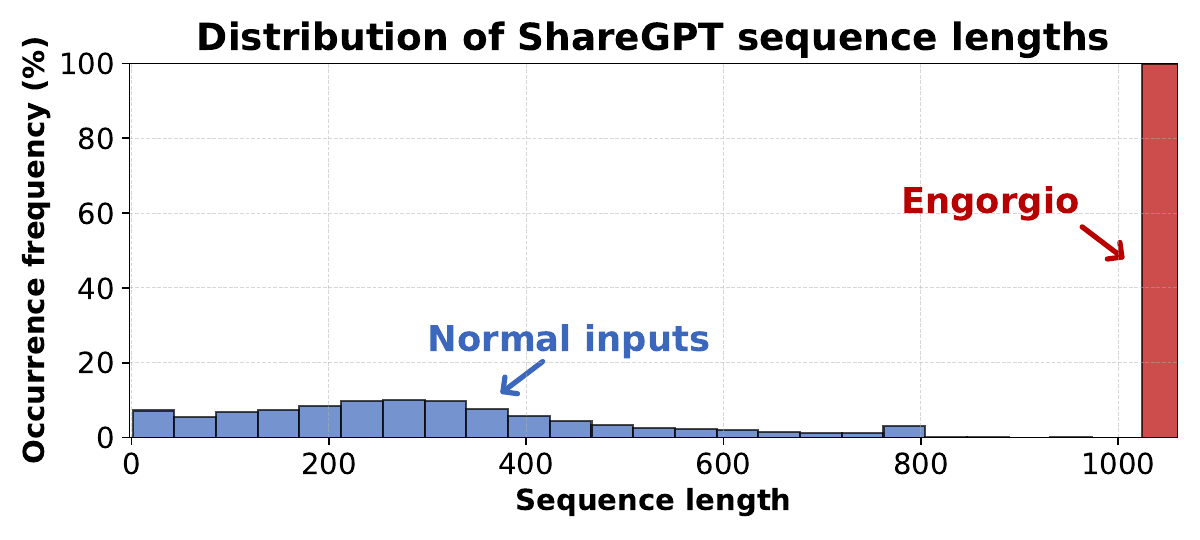}
\vspace{-2em}
%\caption{Distributions of the total lengths (input plus output) of normal samples from ShareGPT~\citep{sharegpt}$^2$ and Engorgio prompts.}
\caption[Caption for LOF]{Distributions of the total lengths (input plus output) of normal samples from ShareGPT\footnotemark[2] and Engorgio prompts.}
\label{fig:sharegpt-length-dist}
\vspace{-1em}
\end{wrapfigure}
In this paper, we introduce Engorgio\footnote{Engorgio is a spell in the Harry Potter universe, which causes objects (or creatures) to increase in size.}, a simple yet effective method to generate threatening Engorgio prompts against state-of-the-art LLMs. 
Our focus is on text completion, where LLMs predict the next token based on the initial prompt and previously generated tokens until an end-of-sequence (\eos) token is predicted or a maximum length is reached.
Technically, Engorgio effectively addresses the above challenges via:
\textbf{(1)} Inspired by the special role of \eos token in determining whether the model halts its response, we adopt an untargeted objective called \eos escape loss, which reduces the \eos token's occurrence probability. 
We also combine a self-mentor loss to stably induce longer responses.
\textbf{(2)} We employ a re-parameterization design to effectively utilize the gradients, by modeling the potential distribution of the entire context that can fulfill both objectives.
Figure~\ref{fig:sharegpt-length-dist} shows the effects of our attack: normal prompts (e.g., the renowned ShareGPT dataset\footnote{\url{https://sharegpt.com}}) typically tempt the LLMs to produce short sequences; in contrast, the crafted Engorgio prompts can make the model response extraordinarily long. 

In summary, our main contributions lie in three folds:
1) We explore a novel research direction, inference cost attack against modern auto-regressive LLMs.
We highlight how crafted adversarial prompts can impact LLM service availability.
2) We analyze technical challenges associated with the attack surface.
Based on our insights, we propose Engorgio, a simple yet effective method that can stably induce lengthy LLM responses.
3) To prove the effectiveness of Engorgio, we conduct extensive experiments over 6 base models and 7 supervised fine-tuned (SFT) models with parameters ranging from 125M to 30B, as listed in Table~\ref{tab:M}. 
Specifically, the generated Engorgio prompts can achieve roughly 90\%+ of the maximum allowable length on 6 base models, while normal queries can only cause between 0-40\%. For SFT models, Engorgio can significantly outperform baselines by up to 13$\times$. A real-world experiment demonstrates Engorgio's implications in service availability.

% \vspace{-1em}

\section{Preliminaries}

\subsection{Large Language Models (LLMs)} 
\label{bg:lmcost}

The task of language modeling tracks the rationality of text sequences and treats the probability of a certain sequence as a product of conditional probabilities~\citep{jelinek1980interpolated,bengio2003neural}:
\begin{equation}
\begin{aligned} \label{eq:casual}
P(x_1,\cdots,x_N)=\prod_{i=1}^{N}P(x_i|x_1,\cdots,x_{i-1}),
\end{aligned}
\end{equation}
where $P(x_i|x_1,\cdots,x_{i-1})$ denotes the probability of predicting $x_i$ as the next token given the sequence $x_1 \cdots x_{i-1}$. 
For a Transformer-based model $f_{\Theta}:\mathcal{X}\to \mathcal{Y}$, it accepts a sequence of tokens with any admitted length $S$ and produces an output vector $r_S = f_{\Theta}(X_{1:S})\in \mathbb{R}^{V}$ to predict the next token, where $V$ is the model's vocabulary size. 
Most prevalent LLMs like LLaMA~\citep{touvron2023llama} and GPT-4~\citep{openai2023gpt4} are based on the Transformer decoder architecture~\citep{attentionisallyouneed}. 
The architecture is designed to perform inference in an auto-regressive manner~\citep{autoregressive}, i.e., LLMs generate one token at a time and use the previously generated tokens to predict next tokens. 
We detail the LLM generation process and models involved in this work in Appendix~\ref{appx:models}.

\subsection{Inference Cost Attacks}
\label{sec:relatedWorks}

Machine learning services are facing an availability threat. 
\citet{shumailov2021sponge} showed that malicious users could intentionally craft adversarial inputs, known as sponge examples, to significantly increase the energy consumption and latency of the corresponding inference process. 
Such inference cost attacks can greatly affect the service provider's operational cost and user experience. 
Following this work, a variety of attacks have been designed to target different AI systems and applications, for example, image classification \citep{muller2024impact},  camera-based object detection \citep{shapira2023phantom, schoof2024beyond, shapira2022denial, xiao2024sponge, ma2024slowtrack}, LiDAR-based object detection \citep{liu2023slowlidar}, and multimodal models \citep{gao2024inducing}.

This paper focuses on attacking the modern auto-regressive LLMs.
Existing inference cost attacks against language models \citep{shumailov2021sponge,chen2022nmtsloth,feng2024llmeffichecker} are only effective when targeting the encoder-decoder structure.
\citet{shumailov2021sponge} generated sponge examples by compressing more tokens into one sentence, leading to a higher computational burden in the cross-attention operations, which are ineffective for LLMs lacking cross-attention modules.
Sub-word tokenization methods BPE~\citep{sennrich-etal-2016-neural-bpe} eliminate the appearance of \unk token and enhance LLMs' typo-tolerating ability, largely invalidating perturbation-based methods like LLMEffiChecker~\citep{feng2024llmeffichecker}.
For the optimization-based method, \citet{geiping2024coercing} proposes a targeted attack that coerces LLMs into producing elicit a specific starting response (i.e., repeating "Hello There" 24 times), indirectly achieving effects similar to an inference cost attack. 
However, this approach is less stable due to its reliance on the starting response's effectiveness and is easily detectable as the starting response serves as a clear indicator of adversarial intent.
This motivates us to design a new attack methodology tailored for modern LLMs. 
In this work, we propose a simple yet effective method to overcome the technical challenges inherent in this task.

\subsection{Threat Model}
\label{subsec:threatmodels}

We design our attack following the threat model of previous inference cost attack studies against language models~\citep{shumailov2021sponge,chen2022nmtsloth,feng2024llmeffichecker} and provide detailed discussion about the practicality and implications of the attack in Appendix~\ref{appx:feasibility-discussion}.

\begin{packeditemize}
\item  \textbf{Attacker's goal}: As a service user, the attacker aims to craft Engorgio prompts $\mathcal{T}$, which could induce as long output as possible. 
Such behaviors could bring much higher operational costs for the LLM service provider, and affect the service availability to other normal users.

\item \textbf{Attacker's knowledge}: We mainly consider a white-box scenario, where the attacker has full knowledge of the target model, including its architecture, input template, model parameters, etc. 
We also consider the black-box setting, in which we transfer Engorgio prompts to attacker-unknown models (see Appendix~\ref{appx:transferability} for details).

\item \textbf{Attacker's capability}: The attacker locally generates the Engorgio prompts $\mathcal{T}$, aligned with her knowledge settings. Then she sends the constructed Engorgio prompts $\mathcal{T}$ to the target LLMs and collects the responses for attack checking. 
\end{packeditemize}

\section{Methodology}

\subsection{Attack Insight and Overview}

In order to achieve the attack goal, we review the mechanism of generating texts by LLMs. 
A sample for the LLM can be split into an input part and an output part (see Appendix~\ref{appx:context-composition} for more analysis). 
Given an input sequence (dubbed prompt) composed of $k$ tokens, the model generates the subsequent tokens (i.e., the output part).
The generation continues until either of two conditions is met: 
(1) reaching a pre-set maximum allowable length; 
(2) encountering an \eos token which indicates the end of the sentence.
As the maximum allowable length is fixed as $S$, the problem is stated as follows: \textit{the later an} \eos \textit{token is encountered in the inference process, the higher cost and latency this query will take}.
Therefore, to achieve latency damages to the service provider, i.e., maximizing the length of the output part, the attacker aims to create Engorgio prompts,
which can effectively \textit{suppress the possibility of predicting the} \eos \textit{token during the inference}.

Based on this insight, we design Engorgio, a novel attack framework to generate Engorgio prompts against LLMs. Figure~\ref{fig:overview} shows its overall pipeline and we provide a term list in Appendix~\ref{appx:term-list}. The core is the introduction of a parameterized proxy distribution. To satisfy the requirements for Engorgio prompts, we explore how to update the distribution with the guidance of an \eos escape loss and self-mentor loss.
The whole process of crafting Engorgio prompts is two-stage:
\begin{packeditemize}
\item \textbf{Generation stage}: For each optimization step, we convert the proxy distribution matrix $\theta$ to a weight matrix $w$ using the Gumbel-Softmax function. We then aggregate the embeddings of all token candidates weighted by $w$ to project $\theta$ into the embedding space. This output is fed into the model to calculate two loss terms, allowing us to obtain gradients for $\theta$ easily. The matrix $\theta$ is updated based on these losses, continuing until no significant changes are detected.

\begin{figure*}[t]
    \centering
    \includegraphics[width=0.9\linewidth]{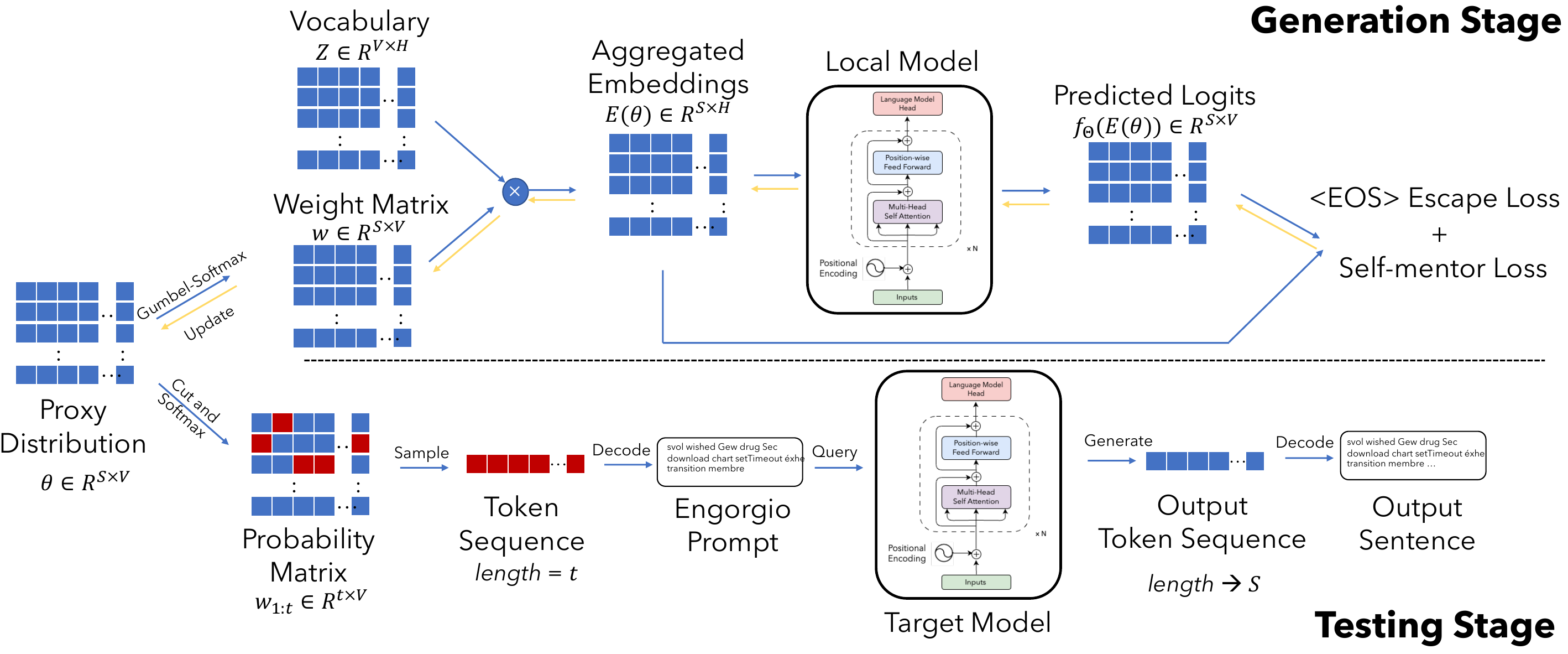}
    \vspace{-1em}
    \caption{The pipeline of Engorgio. 
    The whole pipeline is divided into two stages.
    During the \textbf{generation stage}, we employ a gradient-based method to update the proxy distribution for the Engorgio prompt, where the gradient information is obtained from a local proxy model.
    For the \textbf{testing stage}, we leverage the optimized proxy distribution to decide the final Engorgio prompt.}
    \label{fig:overview}
    \vspace{-1em}
\end{figure*}
 
\item \textbf{Testing stage}:
The optimization process guarantees that the output part falls onto a region with low probabilities of \eos. 
Given the strong correlation between the Engorgio prompt and the output, we can sample the Engorgio prompt using the normalized $w_{1:t}$. 
It is observed that as the optimization progresses, the distribution matrix $\theta$ typically converges toward a specific prompt with a significantly higher sampling probability compared to others. 
This prompt is adopted as the final Engorgio prompt $\mathcal{T}$.
This approach significantly reduces the cost of evaluating other prompt candidates.
We hypothesize that the objectives, particularly self-mentor loss, contribute to identifying the optimal Engorgio prompt, as detailed in Section~\ref{subsec:analysis}.
\end{packeditemize}

\subsection{Proxy Distribution}
To increase the lengths of the target LLM's responses, we search for the Engorgio prompts $\mathcal{T}$ with the help of a proxy model. 
LLMs typically accept a token sequence (corresponding to one input text) as input, cast each token into the embedding space, and work within the embedding space.
Each token has a corresponding embedding; however, not all embeddings correspond to tokens.
We can optimize suitable embedding expressions that satisfy the objectives in the form of prompt learning~\citep{li2021prefix-tuning, ptuning}, but we face challenges in determining the corresponding token sequence (i.e., input text).
We resort to a re-parameterization method.

As LLMs predict the next tokens according to the probability distribution, it is more efficient to search for desirable Engorgio prompts by sampling from an appropriate distribution~\citep{guo-etal-2021-gradient}. 
Therefore, we introduce a proxy distribution to track the process of sequence sampling. 
This proxy distribution is parameterized as a matrix $\theta \in \mathbb{R} ^ {S \times V}$, with $S$ denoting the maximum allowable length, corresponding to the whole context.
It instructs how to select a suitable token sequence from a token vocabulary with $V$ token candidates in the following test stage. 
Then the question is how to ensure that the proxy distribution $\theta$ meets the objectives. 
This endeavor seeks to involve the distribution matrix in the generation stage, subsequently updating it based on the gradients.
Concretely, in the forward pass, the distribution vector $\theta_i$, corresponding to the $i$-th token in the Engorgio prompt where $i \in\left\{ 1,\cdots ,S \right\}$, is independently normalized. 
This serves as a weight to aggregate token embeddings across the model vocabulary, thereby casting $\theta_i$ as a soft token in the embedding space. 
This process is formulated as Eq.~\ref{eq:enc}.
\vspace{-0.5em}
\begin{equation}
\begin{aligned} \label{eq:enc}
\tilde{e}(\theta_i)=\sum_{j=1}^{V}(w_i)_je(j),
\end{aligned}
\end{equation}
where $e(j) \in \mathbb{R}^{H}$ denotes the embedding of the $j$-th token within the model vocabulary and $w_i \in \mathbb{R}^{V}$ is the normalized version of $\theta_i$ with the sum $\sum_{j=1}^{V}(w_i)_j=1$. 
We adopt Gumbel-Softmax~\citep{gumbelsoftmax}, which introduces stochastic elements and enriches the diversity of tokens involved in the generation stage. 
The normalization is conducted in Eq.~\ref{eq:gumbel_softmax}:
\begin{equation}
\begin{aligned} \label{eq:gumbel_softmax}
(w_i)_j = \tfrac{\text{exp}((\theta_{i,j}+g_{i,j})/\tau)}{\sum_{k=1}^{V}\text{exp}((\theta_{i,k}+g_{i,k})/\tau)},
\end{aligned}
\end{equation}
where $g_{i,1} \cdots g_{i,V}$ are drawn from the distribution Gumbel(0,1) and $\tau > 0$ is a temperature factor used to control the uncertainty. 
The introduction of the random variable $g_{i,k}$ from an i.i.d distribution benefits the diversity of the sampling operation.
Due to the differentiability of Gumebl-Softmax, we can take full advantage of the gradient information to update $\theta$ in the generation stage and guide the sampling of the final Engorgio prompt $\mathcal{T}$ in the test stage. 

SFT models assume the input should be embedded in a specified template $T$, as illustrated in Figure~\ref{fig:seq-composition}. 
Considering the most general case that the template $T = \left\{ \left[ P_{1:i} \right],x,\left[ P_{i+1:m} \right],y \right\}$ contains a prefix and an infix, we define the corresponding embedding sequence to $\theta$ as Eq.~\ref{eq:input_form}.
\begin{equation}
\begin{aligned} \label{eq:input_form}
E(\theta)= \left\{ e(\left[ P_{1:i} \right]),\tilde{e}(\theta_{1:t}),e(\left[ P_{i+1:m} \right]),\tilde{e}(\theta_{t+1:s-m}) \right\},
\end{aligned}
\end{equation}
where $P_{1:i}$ and $P_{i+1,m}$ represent the token sequences corresponding to prefix and infix, and the shape of $\theta$ is adjusted to $(s-m)\times V$. 
The input composition is illustrated in Figure~\ref{fig:seq-composition}.

\subsection{Loss Design} \label{subsec:loss design}
To obtain a desirable proxy distribution, we mainly depend on two key loss components to update $\theta$, \eos escape loss and self-mentor loss. 
The \eos escape loss closely aligns with our target goal to make the output part longer while the self-mentor loss is designed to enhance the usability of the proxy distribution. 
Balancing the impact of the two loss terms with $\lambda$, we update the proxy distribution as follows:
\begin{equation}
\begin{aligned} \label{eq:opt}
\min_{\theta} \ \mathcal{L}_{esc}(\theta)+\lambda\mathcal{L}_{sm}(\theta)
\end{aligned}
\end{equation}
\noindent\textbf{\eos escape loss.} \label{subsubsec:eos loss}
Due to the unpredictability of the LLM generation process, enforcing a specified long response is challenging.
We resort to an untargeted objective, which is to decrease the prediction probability of the \eos token. 
However, it is still impossible to accurately forecast the exact occurrence position of \eos during the test stage.
To tackle this, we propose penalizing the occurrence of \eos token from all positions, rather than focusing on specific positions. 
This broader treatment allows us to effectively manage the uncertainties associated with \eos placement.
The \eos escape loss is defined as below:
\begin{equation}
\begin{aligned} \label{eq:escloss}
\mathcal{L}_{esc}(\theta) =  \sum_{i=1}^{S}\text{Softmax}(f_{\Theta}(E(\theta)_{1:i}))_\kappa,
\end{aligned}
\end{equation}
where $\kappa$ denotes the index of the \eos token for the target model. 
We adopt a Softmax-normalized probability of \eos so that it can better measure the relative chance that the model predicts \eos as the next token at a certain position, which is more effective than directly decreasing the absolute logit of the \eos token.  
An input sequence containing \eos is illegal, as the inference process should have halted before predicting the next tokens for the Engorgio prompt.
Therefore, we also consider reducing the predicted \eos probabilities of the Engorgio prompt part.  

\noindent\textbf{Self-mentor loss.}
Another challenge is that we can only query the target model utilizing the Engorgio prompt $\mathcal{T}$ to ensure attack stealthiness and efficiency.
Considering the auto-regression nature of modern LLMs, we cut off the first $t$ tokens as our Engorgio prompt. 
Moreover, $\theta_i$ only independently tracks the token selection of the $i$-th position, but the correlation between tokens should also be enhanced. 
Therefore, we seek to enhance the relevance of all tokens in the sequence, especially the bond between the Engorgio prompt and output parts. 
Inspired by LLM's causal pre-training paradigm, we search for a sequence where the proxy model fits well. 
The loss term is given below: 
\begin{equation}
\begin{aligned} \label{eq:smloss}
\mathcal{L}_{sm}(\theta) = \sum_{i=1}^{S} \mathcal{L}(w_{i+1},\text{Softmax}(f_{\Theta}(E(\theta)_{1:i}))),
\end{aligned}
\end{equation}
where $\mathcal{L}$ is the cross entropy loss. 
The closer to 0 $\mathcal{L}_{sm}$ is, the better the proxy model fits in input $E(\theta)_{1:S}$, which helps the Engorgio prompt $\mathcal{T}$ steadily induce a longer output.

\section{Evaluation} \label{sec:exp}

\subsection{Experimental Setup}  \label{subsec:exp setup}

\noindent \textbf{LLMs.} We include multiple base models, OPT-125M,  OPT-1.3B, GPT2-large, LLaMA-7B, LLaMA-2-7B, and LLaMA-30B.
SFT models are further fine-tuned with additional datasets, for which we consider seven well-known SFT models including Alpaca (7B), Vicuna (7B), StableLM (7B), Koala (7B), Oraca (7B), Samantha (7B), and ChatGLM (6B), as our targets.
More details about the models involved in this work are provided in Appendix~\ref{appx:models}.
Considering the crucial importance of prompts, we also consider the three cases of deploying base models with prompts according to the attacker's knowledge about the deployed prompt (cf. Appendix~\ref{appx:prompt-cases}).

\noindent \textbf{Baselines.} We consider four types of inputs as baselines for comparisons. 
(1) Normal inputs: we collect 50 samples from the training dataset for Standford-alpaca\footnote{\url{ https://github.com/tatsu-lab/stanford_alpaca/}}, which are generated by OpenAI's text-davinci-003, and 50 samples from ShareGPT\footnote{\url{https://sharegpt.com/}}, a website where people can share their ChatGPT conversations. We use the mixup to roughly represent the normal response length of LLMs.
(2) Special inputs: we use prompts with the semantics of demanding a longer output ($i.e.$, prompts starting with ``output longer''). 
(3) LLMEffiChecker: we adopt the three attacks (character, word, and structure attack) proposed in~\citet{feng2024llmeffichecker} and report the averaged results across the attack variants.
(4) Sponge examples: we generate such inputs using the method from~\citet{shumailov2021sponge} by only setting the same input length as our method.

\noindent \textbf{Metrics.} 
Due to the intractable serving mechanisms for LLM, we report results on the level of model behaviors.
To mitigate potential sampling bias caused by the inherent variability in LLM inference, we measure the average token number of the generated outputs (\textbf{Avg-len}).
We query the target LLM multiple times using the sampling generation and compute the average length across these responses. This renders \textbf{Avg-len} a robust estimate of the Engorgio prompt's efficacy.
Second, we calculate the ratio of the LLM outputs that reach the maximum length (\textbf{Avg-rate}) to evaluate the stability.
Notably, inference costs increase super-linearly with longer responses, making \textbf{Avg-len} a lower bound on the prompt's impact on inference cost, which we detail in Appendix~\ref{appx:further-severity}.

\noindent \textbf{Configurations.} 
We use the Adam optimizer with a learning rate of 0.1 to update the distribution matrix $\theta$. 
We allow a maximum of 300 optimization steps, the cost of which is acceptable, especially when considering the reusability as explained in Appendix~\ref{appx:economical-aspect}.
The Gumbel-Softmax temperature factor $\tau$ is set to 1, and the default Engorgio prompt length is $t=32$.
The input length of normal inputs, special inputs, LLMEffiChecker, and sponge examples is roughly the same as Engorgio to ensure fairness. 
The loss coefficient $\lambda$ is empirically set to 1.
The optimization starts with a random prompt, which we use to initialize the proxy distribution.
Constrained by the computing resources, we set 1,024 as the pre-set maximum length.
We also conduct experiments with full context size on two representative base models (2,048 for LLaMA-30B and LLaMA-7B) and one SFT model (4,096 for Samantha) to demonstrate the extensibility to longer context size.
Please refer to Appendix~\ref{appx:input-examples} for examples of prompts and responses for normal inputs, sponge examples, and Engorgio prompts.

\subsection{Main Results} \label{subsec:main-exp}

We report our results on base models in Table~\ref{tab:main-results}. 
Comparing normal and special inputs reveals that semantic inputs induce base models to output longer. 
This means that base LLMs can understand the semantics inside the inputs and seemingly feature being talkative.
However, relying solely on special input is far from reaching the maximum allowable length. 
LLMEffiChecker proves ineffective against more advanced LLMs.
Our method can achieve a very high ratio (roughly 90-100\%) of letting the base model keep endlessly generating tokens until reaching its maximum length, which outperforms all baselines including sponge examples. 
While sponge examples extend output length compared to normal or special inputs, they are less stable than Engorgio as they struggle with LLMs' sampling-based decoding.
Results of more base models are presented in Appendix~\ref{appx:more-base-models}.

\begin{table*}[t]
\centering
\small
\caption{Results of Engorgio against modern LLMs.}
\label{tab:main-results}
\vspace{-1em}
\resizebox{0.95\linewidth}{!}{
\begin{tabular}{ccccccccc} 

\Xhline{1pt}
\textbf{Model} & \multicolumn{2}{c}{\textbf{LLaMA-30B}} & \multicolumn{2}{c}{\textbf{LLaMA-7B}} & \multicolumn{2}{c}{\textbf{LLaMA-7B}} & \multicolumn{2}{c}{\textbf{LLaMA-2-7B}} \\ 
\textbf{Max length} & \multicolumn{2}{c}{2048} & \multicolumn{2}{c}{1024} & \multicolumn{2}{c}{2048} & \multicolumn{2}{c}{1024} \\ 
 & Avg-len & Avg-rate & Avg-len & Avg-rate & Avg-len & Avg-rate & Avg-len & Avg-rate \\ \Xhline{1pt}
\textbf{Normal inputs} &622.2 &12\% &611.4 & 39\% & 757.9&16\% & 818.9 & 69\% \\
\textbf{Special inputs} &1005.2 &40\% &737.3 &58\% &1292.2 &54\% &773.8 & 50\%  \\
\textbf{LLMEffiChecker} & 1052.8 & 28\% & 682.3 & 41\% & 1306.7 & 41\% & 833.7 & 64\% \\
\textbf{Sponge examples} & 1277.7&38\% &857.6 &81\% &1659.8 &78\% & 900.6 & 86\% \\
\textbf{Engorgio} &\textbf{2019.1} &\textbf{95\%}&\textbf{983.4} &\textbf{94\% }&\textbf{1817.7} &\textbf{84\%} & \textbf{1024} & \textbf{100\%}  \\  
\Xhline{1pt}

\textbf{Model} & \multicolumn{2}{c}{\textbf{Samantha}}  & \multicolumn{2}{c}{\textbf{StableLM}} & \multicolumn{2}{c}{\textbf{Koala}}  &\multicolumn{2}{c}{\textbf{Orca}} \\ 
\textbf{Max length} & \multicolumn{2}{c}{4096} & \multicolumn{2}{c}{1024} &\multicolumn{2}{c}{1024} &\multicolumn{2}{c}{1024} \\
& Avg-len & Avg-rate & Avg-len & Avg-rate & Avg-len & Avg-rate & Avg-len & Avg-rate \\ \Xhline{1pt}
\textbf{Normal inputs} & 313.7 & 2\% & 388.2 & 6\% & 357.5 & 6\% & 286.0 & 1\% \\
\textbf{Special inputs} & 173.3 & 0\% & 202.9 & 4\% & 436.1 & 0\% & 199.5 & 0\% \\
\textbf{LLMEffiChecker} & 172.6 & 0\% & 688.3 & 38\%& 324.6 & 5\% & 203.4 & 0\% \\
\textbf{Sponge examples} & 284.9 & 3\% & 301.1 & 16\% & 432.2 & 22\% & 211.1 & 1\% \\
\textbf{Engorgio} & 3951.5 & \textbf{95\%} & 1021.6 & 98\% & \textbf{1024} & \textbf{100\%} & 908.1 & 86\% \\
\textbf{Prefix+Engorgio} & \textbf{4027.6} & \textbf{95\%} & \textbf{1024} & \textbf{100\%} & \textbf{1024} & \textbf{100\%} & \textbf{962.6} & \textbf{90\%} \\ 

\Xhline{1pt}
\textbf{Model} & \multicolumn{2}{c}{\textbf{Samantha}}  & \multicolumn{2}{c}{\textbf{ChatGLM}} & \multicolumn{2}{c}{\textbf{Alpaca}}  &\multicolumn{2}{c}{\textbf{Vicuna}} \\ 
\textbf{Max length} & \multicolumn{2}{c}{1024} & \multicolumn{2}{c}{1024} &\multicolumn{2}{c}{1024} &\multicolumn{2}{c}{1024} \\
& Avg-len & Avg-rate & Avg-len & Avg-rate & Avg-len & Avg-rate & Avg-len & Avg-rate \\ \Xhline{1pt}
\textbf{Normal inputs} & 231.6 & 2\% & 263.4 & 1\% & 179.4 & 0\% & 312.6 & 0\% \\
\textbf{Special inputs} & 82.0 & 0\% & 247.9 & 0\% & 132.5 & 7\% & 252.4 & 4\% \\
\textbf{LLMEffiChecker} & 149.8 & 0\% & 182.0 & 0\%& 192.9 &5\% &  273.4 & 0\% \\
\textbf{Sponge examples} & 155.1 & 0\% & 685.2 & 56\% & 833.8 & 78\% & 599.6 & 44\% \\
\textbf{Engorgio} & 944.0 & \textbf{89\%} & 979.6 & 95\% & 954.2 & 92\% & 789.3 & 60\% \\
\textbf{Prefix+Engorgio} & \textbf{970.0} & \textbf{89\%} & \textbf{1024} & \textbf{100\%} & \textbf{1024} & \textbf{100\%} & \textbf{861.5} & \textbf{68\%} \\ 
\Xhline{1pt}
\end{tabular}
}
\vspace{-1em}
\end{table*}

SFT models may use cut-off as a preprocessing strategy on their fine-tuning datasets (e.g., at most 512 tokens for Alpaca).
This potentially biases the fine-tuned model to produce short responses, which makes our goal challenging, as suggested by the results of normal inputs in Table~\ref{tab:main-results}.
For special inputs, even with instructions for longer responses, SFT models still produce notably shorter outputs, sometimes even shorter than normal inputs.
The silent nature of SFT models worsens the performance of sponge examples.
For LLMEffiChecker, the weaker performance extends to SFT models.
We hypothesize that recent LLMs are more robust to typing errors, invalidating perturbation-based attacks.
In contrast, Engorgio knows how to better optimize the Engorgio prompt by focusing on a distinct goal: avoiding the generation of the \eos token.
It effectively increases the output length to approach the maximum limit, especially when paired with a semantic prefix as discussed in Section~\ref{subsec:temp}, achieving near-maximum allowable lengths.
We also explore a black-box scenario, where we resort to the transferability of Engorgio prompts (See Appendix~\ref{appx:transferability}) for details.

\subsection{Ablation Study}
\label{subsec:ablation study}

\noindent\textbf{Impact of loss design.}
% We explore the efficacy of the two loss terms. 
Initially, we assess the performance when optimizing only with the \eos escape loss (noted as ``ESC'' in Table \ref{tab:ablation study}). 
A comparison with normal input and special input from Table \ref{tab:main-results} reveals that even utilizing only the \eos escape loss consistently results in longer outputs. 
We also observe that combining the self-mentor loss, identified as ``ESC+Self-mentor'' in Table \ref{tab:ablation study}, further increases the Avg-len with almost no extra cost.
More experiments in Appendix~\ref{appx:loss-coefficient} show that Engorgio is not strongly dependent on the choice of $\lambda$.

\noindent\textbf{Impact of Engorgio prompt length.}
We explore the attack results under different prompt lengths $t$. 
The basic intuition is that a longer Engorgio prompt can contain more malicious information to induce the LLMs' outputs to be longer. 
The results are given in Table \ref{tab:ablation study} with three different prompt lengths (i.e., 32, 64, and 128).
For base models like OPT-125M and LLaMA-7B, even the smallest prompt length of 32 can induce them to output max-length sequences.
For SFT models, as the prompt length increases, Avg-len and Avg-rate increase in most cases. 
In summary, although a longer prompt improves the attack performance, it is not a prerequisite for ensuring effectiveness.

\begin{table}[!t]
\centering
\begin{minipage}[t]{0.65\textwidth}
  \caption{Ablation study. Prompt length is separated from Avg-len to better understand the impact of key designs.}
  \label{tab:ablation study}
  \vspace{-1em}
 \scalebox{0.95}{\resizebox{\textwidth}{!}{
\begin{tabular}{c|C{2cm}C{1.5cm}C{2cm}C{1.5cm}}
  		\Xhline{1pt}
  		&\multicolumn{4}{c}{\textbf{LLaMA-7B (1024)}}   \\
		\multirow{1.25}{*}{\shortstack{\textbf{Prompt length}}}
          & \multicolumn{2}{c}{ \textbf{ESC}} &\multicolumn{2}{c}{\textbf{ESC+Self-mentor}}\\ 
		 &Avg-len &Avg-rate &Avg-len &Avg-rate\\
  		\Xhline{1pt}
            32   & 893.5 + 32 &80\%  & \textbf{951.4} + 32& 94\% \\
            64 &  870.6 + 64 & 85\%  &  945.5 + 64& 98\% \\
            128   &827.8 + 128& 85\% &880.6 + 128 & 94\%  \\
  		\Xhline{1pt}
		&\multicolumn{4}{c}{\textbf{Alpaca (1024)}}   \\
		\multirow{1.25}{*}{\shortstack{\textbf{Prompt length}}}
          & \multicolumn{2}{c}{ \textbf{ESC}} &\multicolumn{2}{c}{\textbf{ESC+Self-mentor}}\\ 
		 &Avg-len &Avg-rate &Avg-len &Avg-rate\\
  		\Xhline{1pt}
            32&  967.4 + 32 &  96\% & \textbf{992.0} + 32  &  100\%  \\
            64&  943.8 + 64 &  96\% & 949.1 + 64  &  98\%  \\
            128&  867.3 + 128 & 96\%  & 896.0 + 128  & 100\%   \\
  		\Xhline{1pt}
        &\multicolumn{4}{c}{\textbf{Koala (1024)}}   \\
        \multirow{1.25}{*}{\shortstack{\textbf{Prompt length}}}
          & \multicolumn{2}{c}{ \textbf{ESC}} &\multicolumn{2}{c}{\textbf{ESC+Self-mentor}}\\ 
		  &Avg-len &Avg-rate &Avg-len &Avg-rate\\
  		\Xhline{1pt}
            32&  980.0 + 32 &  98\% &  \textbf{992.0} + 32 & 100\% \\
            64& 950.5 + 64  & 98\%  & 960.0 + 64  &  100\%  \\
            128& 849.2 + 128  & 90\%  & 896.0 + 128  &  100\%  \\
            \Xhline{1pt}
	\end{tabular}
	}}

\end{minipage}\hfill
\begin{minipage}[t]{0.32\textwidth}
\centering
\caption{Impact of temperature setting.}
\label{tab:temperature}
\vspace{-1em}
\resizebox{\textwidth}{!}{
\begin{tabular}{c|c|c} \Xhline{1pt}
\multicolumn{3}{c}{\textbf{StableLM (1024)}} \\ 
\textbf{Temperature} & \textbf{Avg-len} & \textbf{Avg-rate}\\ \Xhline{1pt}
0.1 & 1021.6 & 98\% \\ 
0.3 & 830.5 & 62\% \\ 
0.5 & 610.4 & 28\% \\ 
0.7 & 513.8 & 33\% \\ 
\Xhline{1pt}
\multicolumn{3}{c}{\textbf{Samantha (1024)}} \\ 
\textbf{Temperature} & \textbf{Avg-len} & \textbf{Avg-rate}\\ \Xhline{1pt}
0.1 & 944.1 & 90\% \\ 
0.3 & 714.1 & 58.7\% \\ 
0.5 & 553.0 & 40\% \\ 
0.7 & 406.3 & 23.8\% \\ 
\Xhline{1pt}
\multicolumn{3}{c}{\textbf{ChatGLM (1024)}} \\ 
\textbf{Temperature} & \textbf{Avg-len} & \textbf{Avg-rate}\\ \Xhline{1pt}
0.1 & 979.6 & 95\% \\ 
0.3 & 934.0 & 88\% \\ 
0.5 & 908.1 & 81\% \\ 
0.7 & 820.0 & 71\% \\ 
\Xhline{1pt}
\end{tabular}
}
\end{minipage}
\vspace{-1.5em}
\end{table}

\subsection{Attacks at Different Decoding Temperatures} \label{subsec:temp}
We investigate how temperature affects Engorgio's effectiveness.
Results in Table~\ref{tab:temperature} show that a larger temperature introduces more uncertainty during generation, potentially leading to deviations in the model response.
For talkative base models, they are tempted to respond endlessly when a high temperature of 0.7 is used while a low temperature of 0.1 is more suitable for the silent SFT models.
In most cases, e.g., when querying API service, the temperature is at the users' discretion.
Our quantitative statistics show that the output lengths induced by Engorgio prompts gather either at the shorter end or around the maximal length. See details in Appendix~\ref{appx:response-dist}.
Engorgio prompts can either encourage the SFT model to generate longer outputs or confuse it, resulting in brief responses like "not understand."
Thus, we consider fusing Engorgio prompts with semantic instructions.

\begin{wraptable}{r}{0.5\textwidth}
    \centering
    \vspace{-1em}
    \caption{Results of introducing semantic prefix/suffix when the temperature is 0.7.}
    \label{tab:ica-prefix}%
    \vspace{-1em}
    \resizebox{0.5\textwidth}{!}{
    \begin{tabular}{c|cccc} \Xhline{1pt}
     & \multicolumn{2}{c}{\textbf{Alpaca}} & \multicolumn{2}{c}{\textbf{StableLM}} \\
    \textbf{Model}& \multicolumn{2}{c}{1024} &\multicolumn{2}{c}{1024}  \\
     & Avg-len & Avg-rate &Avg-len & Avg-rate  \\ \hline
    \textbf{Only prefix} & 132.5 & 7\% & 202.9 & 4\% \\
    \textbf{Prefix+normal} & 214.2 & 1\%  & 440.6 & 9\% \\
    \textbf{Engorgio} & 353.2 & 8\%& 513.8 & 33\%\\
    \textbf{Prefix + Engorgio} &\textbf{531.2} & \textbf{43\%} & \textbf{884.8} & \textbf{83\%} \\
    \textbf{Engorgio + suffix} & 314.9 & 18\% & 534.3 & 38\%\\
    \Xhline{1pt}
    \end{tabular}
    }
    \vspace{-1em}
\end{wraptable}
\noindent \textbf{Adding semantic prefix/suffix.} 
We can introduce additional semantic instructions to avoid the SFT model directly outputting ``\emph{not understood}''. 
Particularly, a prefix with semantics that encourages longer response will be woven with the Engorgio prompt in both the generation and testing stage. 
The results in Table~\ref{tab:ica-prefix} show that introducing semantic prefixes can improve performance. 
Comparing with the results in Table~\ref{tab:main-results}, we can observe that adding the prefix to normal inputs still cannot induce an extremely long response. 
Compared to adding a prefix, adding the same semantic sequence as a suffix does not help. 
We hypothesize that that's because a semantic prefix impacts the entire generation process while a suffix only influences the subsequent generation.

\subsection{Attacking Real-world LLM Services}
\label{subsec:case-study}

We conduct a real-world case study to assess the practical threats of Engorgio.
Corresponding to realistic scenarios listed in Appendix~\ref{appx:feasibility-discussion}, users share limited cloud resources for inference requests.

\noindent \textbf{Experiment setup.} 
We utilize the Hugging Face inference endpoint\footnote{\url{https://ui.endpoints.Huggingface.co/}} as our cloud service, deploying StableLM (maximal length of 4096) as the target LLM. 
Our experiments explore three GPU configurations: 1× Nvidia A10, 4× Nvidia A10, and 2× Nvidia A100, aiming to demonstrate how a small number of attackers can significantly compromise the service's performance.
We focus on two main metrics: normal client \textit{latency}, defined as the average response time from querying the service to receiving the output, and server \textit{throughput}, calculated as the number of requests processed per minute.
More details can be referred to in Appendix~\ref{appx:real-world-attack-setup}.

\begin{figure}[tbp]
    \centering
    \includegraphics[width=0.9\textwidth]{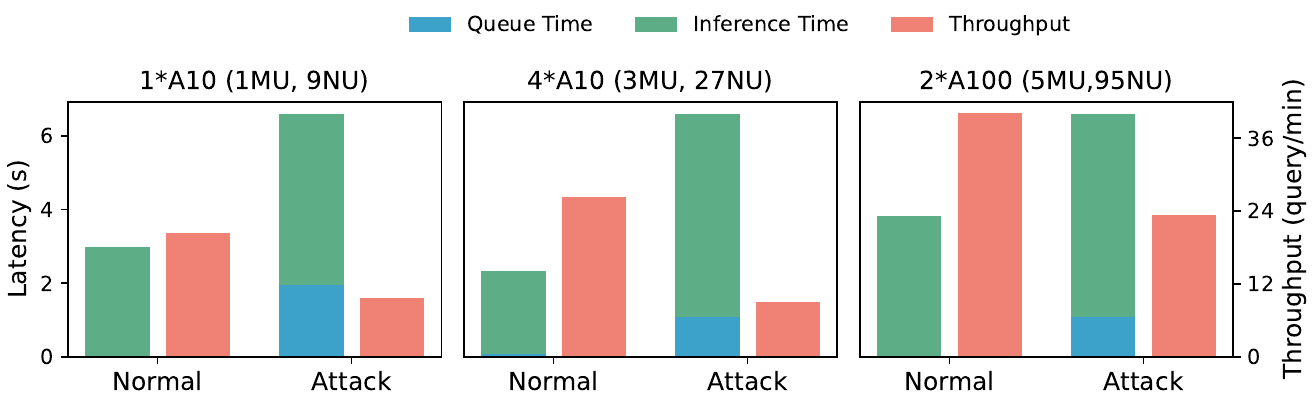}
    \vspace{-1em}
    \caption{Results of attacking real-world LLM services (``MU'': malicious user, ``NU'': normal user).}
    \label{fig:real_word}
\end{figure}

\noindent \textbf{Main results.} 
As shown in~Figure~\ref{fig:real_word}, attackers with Engorgio prompts can significantly compromise the LLM services. 
Although the inference time for normal clients remains consistent, Engorgio prompts significantly increase the queuing time for normal clients scheduled after the attackers.
We observe that only a small ratio of attackers  (e.g., 1 out of 10 or 5 out of 100) could lead to a significant latency increase. 
Besides the negative effect on clients, Engorgio also severely harms the cloud service throughput, which is almost cut off. 
We conclude that a limited number of attackers equipped with Engorgio could severely disturb the fragile cloud-based LLM services.

\subsection{Why is Our Method Effective?} \label{subsec:analysis}

\begin{figure}[tbp]
    \centering
    \vspace{-1em}
    \begin{minipage}[t]{0.56\linewidth}  %
        \centering
        \scalebox{1}{
            \includegraphics[width=\linewidth]{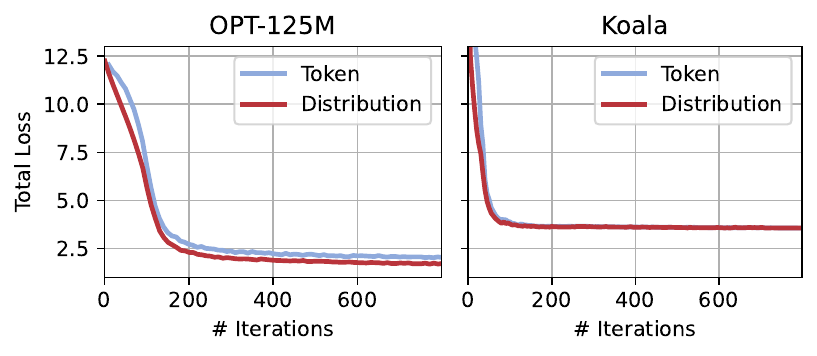}
        }
        \vspace{-2em}
        \caption{Loss curves on OPT-125M (base model) and Koala (SFT model), with aggregated embeddings and token sequence as input, respectively.}
        \label{fig:loss tracking}
    \end{minipage}%
    \hfill
    \begin{minipage}[t]{0.4\linewidth}  %
        \centering
        \scalebox{1}{
            \includegraphics[width=\linewidth]{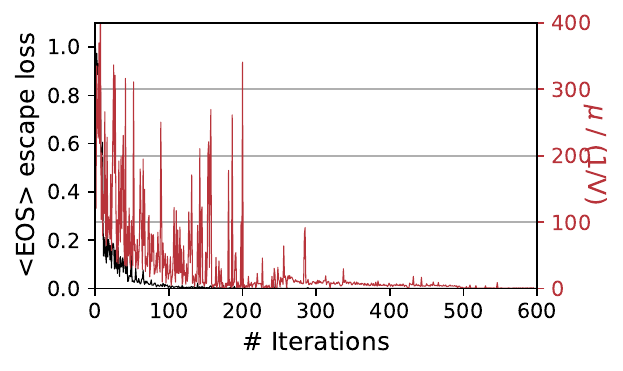}
        }
        \vspace{-2em}
        \caption{On LLaMA-7B, the \eos escape loss correlates with the relative level of \eos being predicted.}
        \label{fig:explainability}
    \end{minipage}
    \vspace{-1em}
\end{figure}

\noindent \textbf{\emph{Q1. How does the distribution matrix instruct the token selection in the optimization process?}} 
In the optimization process, we seek to update the distribution matrix $\theta$ rather than selecting individual tokens, meaning that the most suitable input is the aggregated embedding $\tilde{e}(\theta_i)$. 
In Figure~\ref{fig:loss tracking}, we find that even in the middle of the optimization process, the token sequence greedily sampled according to the distribution matrix performs only slightly worse than the aggregated embeddings when our goal is to decrease the total loss.
This means that our distribution update design is effective in searching for suitable token sequences. 
Moreover, we find that the \eos escape loss of SFT models like Koala is much harder to decrease than base models like OPT-125M. 
This partially supports that base models are easier to induce than SFT models.

\noindent \textbf{\emph{Q2. Does \eos escape loss stop \eos from appearing?}}
To verify that \eos escape loss reduces the probability of \eos appearance, we calculate the highest probability of the \eos token at all the $S$ positions. 
We formulate this as $\mu=\max(\{{\text{Softmax}(f_{\Theta}(E(\theta)_{1:i}))_\kappa }\}_{i=1}^{S})$ which signals the highest probability of the interruption of the generation process. 
We report the relative level of $\mu$ compared to the average probability $1/V$ coupled with the change of \eos escape loss in Figure~\ref{fig:explainability}. 
We find that the decrease of \eos escape loss can lead the maximum probability of the occurrence of \eos token to a low level (close to 0). 
This substantiates the effectiveness of \eos escape loss in stopping \eos token from appearing.

\section{Discussions} 

\noindent \textbf{The resistance to potential countermeasures.}
There are no off-the-shelf defenses tailored for Engorgio yet.
Service providers are faced with a trade-off between detection accuracy and service quality.
Although rare, normal inputs may also lead to a long response.
Engorgio prompts are not crafted to be coherent.
However, our experimental results show that simple methods like a perplexity filter will lead to an unacceptably high false positive rate, significantly degrading user experience.
This is rooted in the variability of legitimate user queries themselves.
What's more, introducing semantic prefixes inevitably improves the coherence of Engorgio prompts, but incurs no performance degradation.
Another potential countermeasure is anomaly detection, monitoring the output length distribution of queries and blocking high-risk users. 
However, the method may face problems of false positives and attackers can strategically adjust behaviors to evade detection.
Please refer to Appendix~\ref{appx:potential-defense} for more related experimental results and discussions.
We will explore effective defense mechanisms in our future work.

\noindent \textbf{Potential limitations.}
Although the white-box setting in this work can already cause far-reaching consequences as explained in Appendix~\ref{appx:feasibility-discussion}, we emphasize the need to systematically study the transferability of Engorgio prompts.
The method efficiency in crafting Engorgio prompts should be further improved.
For the current version, we generate one Engorgio prompt at one time.
We plan to extend to a batch method and study the interoperability among different Engorgio prompts.
To address high-temperature cases, we employ semantic prefixes to mitigate issues. Future work will focus on tracking more active model prediction dynamics to eliminate these challenges. Currently, we do not consider coherence when crafting Engorgio prompts. As coherence enables higher stealthiness of Engorgio prompts, we plan to further explore it in our future work.

\section{Conclusion} 
In this paper, we investigate the inference cost threats to modern auto-regressive language models, which tempt the victim models to produce abnormally long outputs and compromise service availability. 
We introduce Engorgio, a novel attack methodology, to effectively generate Engorgio prompts that can significantly lengthen the model responses. 
Driven by the challenges of uncertain generation process and discrete input modality, our work advances in utilizing proxy distribution and untargeted loss to craft threatening Engorgio prompts.
This is achieved by tracking a parameterized distribution of Engorgio prompts and optimizing it to decrease the occurrence probability of the \eos token.
We validate the effectiveness of Engorgio with extensive experiments on 6 base models and 7 SFT models, considering various prompt scenarios.
By inducing the target LLMs to output until their maximum length limits, we achieve roughly 2-13$\times$ more inference cost per query compared to normal inputs.
We also conduct a real-world case study to demonstrate the practical threat posed by Engorgio to cloud-based LLM services.

\section*{Ethics Statement}

% The demand for computing resources used for LLM inference has drawn attention from both academia and industry.
This paper highlights potential adversarial threats to LLM service availability.
Instead of conducting real-world attacks, this work serves as a clarion call for service providers to consider not only maximizing service latency but also the risks to inference costs posed by malicious users. 
Engorgio offers a method for generating threatening prompts, allowing service providers to stress test their online LLM services effectively.
All experiments adhere to principles of trustworthiness and harmlessness.
Note that real-world attack demos in Section~\ref{subsec:case-study} target only our own LLM service, without impacting others.
Our work utilizes open-source models and datasets, ensuring no privacy violations. 
Our work also does not involve any human subject. 
This work does not raise ethical issues in general.

% Unauthorized model reuse has posed a serious threat to the intellectual property rights (IPRs) of the model developers. Model fingerprinting is a promising solution to detect reused models. In this paper, we propose a new paradigm of model fingerprinting dubbed FIT-Print. Our FIT-Print is purely defensive and does not discover new threats. Moreover, our work utilizes the open-source dataset and does not infringe on the privacy of any individual. Our work also does not involve any human subject. As such, this work does not raise ethical issues in general.

\section*{Reproducibility Statement}

The details of models, hyper-parameter settings, and experimental settings can be found in Section~\ref{subsec:exp setup} and Appendix~\ref{appx:real-world-attack-setup}.
The models involved in this work are all openly accessible.
The codes for reproducing our main evaluation results are provided in the anonymous repository. 
We will release the full codes of our methods upon the acceptance of this paper.

\section*{Acknowledgments}
We would like to thank Yiming Li and the anonymous reviewers for their helpful comments.
This work is supported by the National Science Foundation for Distinguished Young Scholars of China under No. 62425201, and the National Science Foundation China under Grant No. 62132011.

\bibliography{references}

\begin{thebibliography}{49}
\providecommand{\natexlab}[1]{#1}
\providecommand{\url}[1]{\texttt{#1}}
\expandafter\ifx\csname urlstyle\endcsname\relax
  \providecommand{\doi}[1]{doi: #1}\else
  \providecommand{\doi}{doi: \begingroup \urlstyle{rm}\Url}\fi

\bibitem[Aghajanyan et~al.(2021)Aghajanyan, Gupta, and Zettlemoyer]{aghajanyan2021intrinsic}
Armen Aghajanyan, Sonal Gupta, and Luke Zettlemoyer.
\newblock Intrinsic dimensionality explains the effectiveness of language model fine-tuning.
\newblock In \emph{Proceedings of the 59th Annual Meeting of the Association for Computational Linguistics (ACL)}, pp.\  7319--7328, 2021.

\bibitem[Aminabadi et~al.(2022)Aminabadi, Rajbhandari, Awan, Li, Li, Zheng, Ruwase, Smith, Zhang, Rasley, et~al.]{aminabadi2022deepspeed}
Reza~Yazdani Aminabadi, Samyam Rajbhandari, Ammar~Ahmad Awan, Cheng Li, Du~Li, Elton Zheng, Olatunji Ruwase, Shaden Smith, Minjia Zhang, Jeff Rasley, et~al.
\newblock Deepspeed-inference: enabling efficient inference of transformer models at unprecedented scale.
\newblock In \emph{International Conference for High Performance Computing, Networking, Storage and Analysis (SC)}, pp.\  1--15. IEEE, 2022.

\bibitem[Bengio et~al.(2003)Bengio, Ducharme, Vincent, and Jauvin]{bengio2003neural}
Yoshua Bengio, R{\'e}jean Ducharme, Pascal Vincent, and Christian Jauvin.
\newblock A neural probabilistic language model.
\newblock \emph{Journal of Machine Learning Research}, 3:\penalty0 1137--1155, 2003.

\bibitem[Bulatov et~al.(2023)Bulatov, Kuratov, and Burtsev]{bulatov2023scaling}
Aydar Bulatov, Yuri Kuratov, and Mikhail~S Burtsev.
\newblock Scaling transformer to 1{M} tokens and beyond with {RMT}.
\newblock \emph{arXiv preprint arXiv:2304.11062}, 2023.

\bibitem[Carlini et~al.(2021)Carlini, Tramer, Wallace, Jagielski, Herbert-Voss, Lee, Roberts, Brown, Song, Erlingsson, et~al.]{carlini2021extracting}
Nicholas Carlini, Florian Tramer, Eric Wallace, Matthew Jagielski, Ariel Herbert-Voss, Katherine Lee, Adam Roberts, Tom~B Brown, Dawn Song, Ulfar Erlingsson, et~al.
\newblock Extracting training data from large language models.
\newblock In \emph{USENIX Security Symposium}, volume~6, 2021.

\bibitem[Chen et~al.(2022)Chen, Liu, Haque, Song, and Yang]{chen2022nmtsloth}
Simin Chen, Cong Liu, Mirazul Haque, Zihe Song, and Wei Yang.
\newblock Nmtsloth: understanding and testing efficiency degradation of neural machine translation systems.
\newblock In \emph{Proceedings of the 30th ACM Joint European Software Engineering Conference and Symposium on the Foundations of Software Engineering}, pp.\  1148--1160, 2022.

\bibitem[Ding et~al.(2023)Ding, Ma, Dong, Zhang, Huang, Wang, Zheng, and Wei]{ding2023longnet}
Jiayu Ding, Shuming Ma, Li~Dong, Xingxing Zhang, Shaohan Huang, Wenhui Wang, Nanning Zheng, and Furu Wei.
\newblock Longnet: Scaling transformers to 1,000,000,000 tokens.
\newblock \emph{arXiv preprint arXiv:2307.02486}, 2023.

\bibitem[Du et~al.(2022)Du, Qian, Liu, Ding, Qiu, Yang, and Tang]{du2022glm}
Zhengxiao Du, Yujie Qian, Xiao Liu, Ming Ding, Jiezhong Qiu, Zhilin Yang, and Jie Tang.
\newblock Glm: General language model pretraining with autoregressive blank infilling.
\newblock In \emph{Proceedings of the 60th Annual Meeting of the Association for Computational Linguistics (ACL)}, pp.\  320--335, 2022.

\bibitem[Feng et~al.(2024)Feng, Han, Chen, and Yang]{feng2024llmeffichecker}
Xiaoning Feng, Xiaohong Han, Simin Chen, and Wei Yang.
\newblock Llmeffichecker: Understanding and testing efficiency degradation of large language models.
\newblock \emph{ACM Transactions on Software Engineering and Methodology}, 2024.

\bibitem[Gao et~al.(2024)Gao, Bai, Gu, Xia, Torr, Li, and Liu]{gao2024inducing}
Kuofeng Gao, Yang Bai, Jindong Gu, Shu-Tao Xia, Philip Torr, Zhifeng Li, and Wei Liu.
\newblock Inducing high energy-latency of large vision-language models with verbose images.
\newblock \emph{arXiv preprint arXiv:2401.11170}, 2024.

\bibitem[Geiping et~al.(2024)Geiping, Stein, Shu, Saifullah, Wen, and Goldstein]{geiping2024coercing}
Jonas Geiping, Alex Stein, Manli Shu, Khalid Saifullah, Yuxin Wen, and Tom Goldstein.
\newblock Coercing llms to do and reveal (almost) anything.
\newblock In \emph{ICLR 2024 Workshop on Secure and Trustworthy Large Language Models}, 2024.

\bibitem[Gimpel et~al.(2023)Gimpel, Hall, Decker, Eymann, L{\"a}mmermann, M{\"a}dche, R{\"o}glinger, Ruiner, Schoch, Schoop, et~al.]{gimpel2023unlocking}
Henner Gimpel, Kristina Hall, Stefan Decker, Torsten Eymann, Luis L{\"a}mmermann, Alexander M{\"a}dche, Maximilian R{\"o}glinger, Caroline Ruiner, Manfred Schoch, Mareike Schoop, et~al.
\newblock Unlocking the power of generative {AI} models and systems such as {GPT-4} and {ChatGPT} for higher education: A guide for students and lecturers.
\newblock Technical report, Hohenheim Discussion Papers in Business, Economics and Social Sciences, 2023.

\bibitem[Graves(2013)]{autoregressive}
Alex Graves.
\newblock Generating sequences with recurrent neural networks.
\newblock \emph{arXiv preprint arXiv:1308.0850}, 2013.

\bibitem[Guo et~al.(2021)Guo, Sablayrolles, J{\'e}gou, and Kiela]{guo-etal-2021-gradient}
Chuan Guo, Alexandre Sablayrolles, Herv{\'e} J{\'e}gou, and Douwe Kiela.
\newblock Gradient-based adversarial attacks against text transformers.
\newblock In \emph{Proceedings of the 2021 Conference on Empirical Methods in Natural Language Processing (EMNLP)}, pp.\  5747--5757, 2021.

\bibitem[Hoffmann et~al.(2022)Hoffmann, Borgeaud, Mensch, Buchatskaya, Cai, Rutherford, Casas, Hendricks, Welbl, Clark, et~al.]{hoffmann2022training}
Jordan Hoffmann, Sebastian Borgeaud, Arthur Mensch, Elena Buchatskaya, Trevor Cai, Eliza Rutherford, Diego de~Las Casas, Lisa~Anne Hendricks, Johannes Welbl, Aidan Clark, et~al.
\newblock Training compute-optimal large language models.
\newblock \emph{arXiv preprint arXiv:2203.15556}, 2022.

\bibitem[Holtzman et~al.(2019)Holtzman, Buys, Du, Forbes, and Choi]{holtzman2019curious}
Ari Holtzman, Jan Buys, Li~Du, Maxwell Forbes, and Yejin Choi.
\newblock The curious case of neural text degeneration.
\newblock \emph{arXiv preprint arXiv:1904.09751}, 2019.

\bibitem[Jang et~al.(2017)Jang, Gu, and Poole]{gumbelsoftmax}
Eric Jang, Shixiang Gu, and Ben Poole.
\newblock Categorical reparameterization with gumbel-softmax.
\newblock In \emph{International Conference on Learning Representations (ICLR)}, 2017.

\bibitem[Jelinek(1980)]{jelinek1980interpolated}
Frederick Jelinek.
\newblock Interpolated estimation of markov source parameters from sparse data.
\newblock In \emph{Proceeding of the Workshop on Pattern Recognition in Practice}, pp.\  381--397, 1980.

\bibitem[Kaplan et~al.(2020)Kaplan, McCandlish, Henighan, Brown, Chess, Child, Gray, Radford, Wu, and Amodei]{kaplan2020scaling}
Jared Kaplan, Sam McCandlish, Tom Henighan, Tom~B Brown, Benjamin Chess, Rewon Child, Scott Gray, Alec Radford, Jeffrey Wu, and Dario Amodei.
\newblock Scaling laws for neural language models.
\newblock \emph{arXiv preprint arXiv:2001.08361}, 2020.

\bibitem[Lee et~al.(2023)Lee, Hunter, and Ruiz]{lee2023platypus}
Ariel Lee, Cole Hunter, and Nataniel Ruiz.
\newblock Platypus: Quick, cheap, and powerful refinement of {LLM}s.
\newblock In \emph{NeurIPS 2023 Workshop on Instruction Tuning and Instruction Following}, 2023.

\bibitem[Li et~al.(2024)Li, Jiang, Gadepally, and Tiwari]{li2024llm}
Baolin Li, Yankai Jiang, Vijay Gadepally, and Devesh Tiwari.
\newblock {LLM} inference serving: Survey of recent advances and opportunities.
\newblock \emph{arXiv preprint arXiv:2407.12391}, 2024.

\bibitem[Li \& Liang(2021)Li and Liang]{li2021prefix-tuning}
Xiang~Lisa Li and Percy Liang.
\newblock Prefix-tuning: Optimizing continuous prompts for generation.
\newblock In \emph{Proceedings of the 59th Annual Meeting of the Association for Computational Linguistics}, pp.\  4582--4597, 2021.

\bibitem[Liu et~al.(2023)Liu, Wu, Yu, Vorobeychik, and Zhang]{liu2023slowlidar}
Han Liu, Yuhao Wu, Zhiyuan Yu, Yevgeniy Vorobeychik, and Ning Zhang.
\newblock Slowlidar: Increasing the latency of lidar-based detection using adversarial examples.
\newblock In \emph{Proceedings of the IEEE/CVF Conference on Computer Vision and Pattern Recognition (CVPR)}, pp.\  5146--5155, 2023.

\bibitem[Liu et~al.(2021)Liu, Zheng, Du, Ding, Qian, Yang, and Tang]{ptuning}
Xiao Liu, Yanan Zheng, Zhengxiao Du, Ming Ding, Yujie Qian, Zhilin Yang, and Jie Tang.
\newblock Gpt understands, too.
\newblock \emph{arXiv preprint arXiv:2103.10385}, 2021.

\bibitem[Ma et~al.(2024)Ma, Wang, Chen, and Shen]{ma2024slowtrack}
Chen Ma, Ningfei Wang, Qi~Alfred Chen, and Chao Shen.
\newblock Slowtrack: Increasing the latency of camera-based perception in autonomous driving using adversarial examples.
\newblock In \emph{Proceedings of the AAAI Conference on Artificial Intelligence (AAAI)}, volume~38, pp.\  4062--4070, 2024.

\bibitem[M{\"u}ller \& Quiring(2024)M{\"u}ller and Quiring]{muller2024impact}
Andreas M{\"u}ller and Erwin Quiring.
\newblock The impact of uniform inputs on activation sparsity and energy-latency attacks in computer vision.
\newblock \emph{arXiv preprint arXiv:2403.18587}, 2024.

\bibitem[OpenAI(2023)]{openai2023gpt4}
OpenAI.
\newblock {GPT-4 Technical Report}, 2023.

\bibitem[Ouyang et~al.(2022)Ouyang, Wu, Jiang, Almeida, Wainwright, Mishkin, Zhang, Agarwal, Slama, Ray, et~al.]{ouyang2022training}
Long Ouyang, Jeffrey Wu, Xu~Jiang, Diogo Almeida, Carroll Wainwright, Pamela Mishkin, Chong Zhang, Sandhini Agarwal, Katarina Slama, Alex Ray, et~al.
\newblock Training language models to follow instructions with human feedback.
\newblock \emph{Advances in Neural Information Processing Systems (NeurIPS)}, 35:\penalty0 27730--27744, 2022.

\bibitem[Patel et~al.(2024)Patel, Choukse, Zhang, Shah, Goiri, Maleki, and Bianchini]{patel2024splitwise}
Pratyush Patel, Esha Choukse, Chaojie Zhang, Aashaka Shah, {\'I}{\~n}igo Goiri, Saeed Maleki, and Ricardo Bianchini.
\newblock Splitwise: Efficient generative llm inference using phase splitting.
\newblock In \emph{51st Annual International Symposium on Computer Architecture (ISCA)}, pp.\  118--132. IEEE, 2024.

\bibitem[Patterson et~al.(2022)Patterson, Gonzalez, H{\"o}lzle, Le, Liang, Munguia, Rothchild, So, Texier, and Dean]{patterson2022carbon}
David Patterson, Joseph Gonzalez, Urs H{\"o}lzle, Quoc Le, Chen Liang, Lluis-Miquel Munguia, Daniel Rothchild, David~R So, Maud Texier, and Jeff Dean.
\newblock The carbon footprint of machine learning training will plateau, then shrink.
\newblock \emph{Computer}, 55\penalty0 (7):\penalty0 18--28, 2022.

\bibitem[Peng et~al.(2023)Peng, Li, He, Galley, and Gao]{peng2023instruction}
Baolin Peng, Chunyuan Li, Pengcheng He, Michel Galley, and Jianfeng Gao.
\newblock Instruction tuning with gpt-4.
\newblock \emph{arXiv preprint arXiv:2304.03277}, 2023.

\bibitem[Radford et~al.(2018)Radford, Narasimhan, Salimans, Sutskever, et~al.]{gpt1}
Alec Radford, Karthik Narasimhan, Tim Salimans, Ilya Sutskever, et~al.
\newblock Improving language understanding by generative pre-training.
\newblock 2018.

\bibitem[Santurkar et~al.(2023)Santurkar, Durmus, Ladhak, Lee, Liang, and Hashimoto]{santurkar2023whose}
Shibani Santurkar, Esin Durmus, Faisal Ladhak, Cinoo Lee, Percy Liang, and Tatsunori Hashimoto.
\newblock Whose opinions do language models reflect?
\newblock In \emph{International Conference on Machine Learning (ICML)}, pp.\  29971--30004. PMLR, 2023.

\bibitem[Schoof et~al.(2024)Schoof, Koffas, Conti, and Picek]{schoof2024beyond}
Coen Schoof, Stefanos Koffas, Mauro Conti, and Stjepan Picek.
\newblock Beyond phantomsponges: Enhancing sponge attack on object detection models.
\newblock In \emph{Proceedings of the 2024 ACM Workshop on Wireless Security and Machine Learning}, pp.\  14--19, 2024.

\bibitem[Sennrich et~al.(2016)Sennrich, Haddow, and Birch]{sennrich-etal-2016-neural-bpe}
Rico Sennrich, Barry Haddow, and Alexandra Birch.
\newblock Neural machine translation of rare words with subword units.
\newblock In \emph{Proceedings of the 54th Annual Meeting of the Association for Computational Linguistics (ACL))}, pp.\  1715--1725, 2016.

\bibitem[Shapira et~al.(2022)Shapira, Zolfi, Demetrio, Biggio, and Shabtai]{shapira2022denial}
Avishag Shapira, Alon Zolfi, Luca Demetrio, Battista Biggio, and Asaf Shabtai.
\newblock Denial-of-service attack on object detection model using universal adversarial perturbation.
\newblock 2022.

\bibitem[Shapira et~al.(2023)Shapira, Zolfi, Demetrio, Biggio, and Shabtai]{shapira2023phantom}
Avishag Shapira, Alon Zolfi, Luca Demetrio, Battista Biggio, and Asaf Shabtai.
\newblock Phantom sponges: Exploiting non-maximum suppression to attack deep object detectors.
\newblock In \emph{IEEE/CVF Winter Conference on Applications of Computer Vision (WACV)}, pp.\  4571--4580, 2023.

\bibitem[Shen et~al.(2023)Shen, Chen, Backes, and Zhang]{shen2023chatgpt}
Xinyue Shen, Zeyuan Chen, Michael Backes, and Yang Zhang.
\newblock In chatgpt we trust? measuring and characterizing the reliability of chatgpt.
\newblock \emph{arXiv preprint arXiv:2304.08979}, 2023.

\bibitem[Shumailov et~al.(2021)Shumailov, Zhao, Bates, Papernot, Mullins, and Anderson]{shumailov2021sponge}
Ilia Shumailov, Yiren Zhao, Daniel Bates, Nicolas Papernot, Robert Mullins, and Ross Anderson.
\newblock Sponge examples: Energy-latency attacks on neural networks.
\newblock In \emph{2021 IEEE European Symposium on Security and Privacy (EuroS\&P)}, pp.\  212--231. IEEE, 2021.

\bibitem[Su et~al.(2022)Su, Lan, Wang, Yogatama, Kong, and Collier]{su2022contrastive}
Yixuan Su, Tian Lan, Yan Wang, Dani Yogatama, Lingpeng Kong, and Nigel Collier.
\newblock A contrastive framework for neural text generation.
\newblock \emph{arXiv preprint arXiv:2202.06417}, 2022.

\bibitem[Touvron et~al.(2023)Touvron, Lavril, Izacard, Martinet, Lachaux, Lacroix, Rozi{\`e}re, Goyal, Hambro, Azhar, et~al.]{touvron2023llama}
Hugo Touvron, Thibaut Lavril, Gautier Izacard, Xavier Martinet, Marie-Anne Lachaux, Timoth{\'e}e Lacroix, Baptiste Rozi{\`e}re, Naman Goyal, Eric Hambro, Faisal Azhar, et~al.
\newblock Llama: Open and efficient foundation language models.
\newblock \emph{arXiv preprint arXiv:2302.13971}, 2023.

\bibitem[Vaswani et~al.(2017)Vaswani, Shazeer, Parmar, Uszkoreit, Jones, Gomez, Kaiser, and Polosukhin]{attentionisallyouneed}
Ashish Vaswani, Noam Shazeer, Niki Parmar, Jakob Uszkoreit, Llion Jones, Aidan~N Gomez, {\L}ukasz Kaiser, and Illia Polosukhin.
\newblock Attention is all you need.
\newblock \emph{Advances in Neural Information Processing Systems (NeurIPS)}, 2017.

\bibitem[Wang et~al.(2022)Wang, Kordi, Mishra, Liu, Smith, Khashabi, and Hajishirzi]{wang2022self}
Yizhong Wang, Yeganeh Kordi, Swaroop Mishra, Alisa Liu, Noah~A Smith, Daniel Khashabi, and Hannaneh Hajishirzi.
\newblock Self-instruct: Aligning language model with self generated instructions.
\newblock \emph{arXiv preprint arXiv:2212.10560}, 2022.

\bibitem[Wu et~al.(2023)Wu, Irsoy, Lu, Dabravolski, Dredze, Gehrmann, Kambadur, Rosenberg, and Mann]{wu2023bloomberggpt}
Shijie Wu, Ozan Irsoy, Steven Lu, Vadim Dabravolski, Mark Dredze, Sebastian Gehrmann, Prabhanjan Kambadur, David Rosenberg, and Gideon Mann.
\newblock Bloomberg{GPT}: A large language model for finance.
\newblock \emph{arXiv preprint arXiv:2303.17564}, 2023.

\bibitem[Xiao et~al.(2024)Xiao, Ma, Yi, and Chen]{xiao2024sponge}
Yong Xiao, Jin Ma, Ping Yi, and Xiuzhen Chen.
\newblock Sponge backdoor attack: Increasing the latency of object detection exploiting non-maximum suppression.
\newblock In \emph{2024 International Joint Conference on Neural Networks (IJCNN)}, pp.\  1--8. IEEE, 2024.

\bibitem[Yao et~al.(2024)Yao, Lou, Qin, and Ren]{yao2024promptcare}
Hongwei Yao, Jian Lou, Zhan Qin, and Kui Ren.
\newblock Promptcare: Prompt copyright protection by watermark injection and verification.
\newblock In \emph{2024 IEEE Symposium on Security and Privacy (S\&P)}, pp.\  845--861. IEEE, 2024.

\bibitem[Zhao et~al.(2023)Zhao, Zhou, Li, Tang, Wang, Hou, Min, Zhang, Zhang, Dong, et~al.]{zhao2023survey}
Wayne~Xin Zhao, Kun Zhou, Junyi Li, Tianyi Tang, Xiaolei Wang, Yupeng Hou, Yingqian Min, Beichen Zhang, Junjie Zhang, Zican Dong, et~al.
\newblock A survey of large language models.
\newblock \emph{arXiv preprint arXiv:2303.18223}, 2023.

\bibitem[Zhou et~al.(2023)Zhou, Liu, Xu, Iyer, Sun, Mao, Ma, Efrat, Yu, Yu, et~al.]{zhou2023lima}
Chunting Zhou, Pengfei Liu, Puxin Xu, Srini Iyer, Jiao Sun, Yuning Mao, Xuezhe Ma, Avia Efrat, Ping Yu, Lili Yu, et~al.
\newblock Lima: Less is more for alignment.
\newblock \emph{arXiv preprint arXiv:2305.11206}, 2023.

\bibitem[Zou et~al.(2023)Zou, Wang, Carlini, Nasr, Kolter, and Fredrikson]{zou2023gcg}
Andy Zou, Zifan Wang, Nicholas Carlini, Milad Nasr, J~Zico Kolter, and Matt Fredrikson.
\newblock Universal and transferable adversarial attacks on aligned language models.
\newblock \emph{arXiv preprint arXiv:2307.15043}, 2023.

\end{thebibliography}
\bibliographystyle{iclr2025_conference}

\clearpage

\appendix

\section{Further statements}

\subsection{Involved models}\label{appx:models}

Mainstream LLMs can be categorized into two main classes as in Table \ref{tab:M} including pre-trained base models and supervised fine-tuned (SFT) models. 
Base models are pre-trained on large-scale unlabelled training corpora in the manner of self-supervised learning like next token prediction~\citep{gpt1} and auto-regressive blank infilling~\citep{du2022glm}.
This process endows the base models with basic language abilities. 
Base models can be fine-tuned~\citep{ouyang2022training} or distilled from a more powerful oracle model~\citep{wang2022self,peng2023instruction}. 
Such SFT models can typically perform better on downstream tasks.
Besides, low overhead to obtain a usable model makes SFT mainstream in the field of LLM development~\citep{zhou2023lima, lee2023platypus}. 

During inference, LLMs iteratively repeat the process of predicting the next tokens until the maximum length limit is reached or an end-of-sequence token (\texttt{<EOS>}) is encountered. 
LLMs can parallelly process all sub-sequences $\left\{ X_{1:i} \right\}_{i=1}^S$ of the whole input in one single forward pass utilizing the mask design and finally outputs $R \in \mathbb{R}^{S\times V}$, where $R_i = f_{\Theta}(X_{1:i})$. 
A new token will be selected according to $R_S$ and its embedding will be concatenated with the previous sequence to form a new sequence $X_{1:S+1}$, used to predict the following tokens. 
Another representative line of LLMs, ChatGLM~\citep{du2022glm}, incorporates a different attention mask but still involves an auto-regressive inference scheme. 
Both types of auto-regressive LLMs are explored in this paper.

Text decoding methods, which decide how to utilize $R_S$ (the predicted next-token logits) to choose a new token, are essential to natural language generation. 
Greedy search~\citep{su2022contrastive} is the simplest way, which directly selects the token with the maximum probability in $R_S$.
In probabilistic sampling, the decoding process replaces the word with the highest probability with probability-based sampling. 
The sampling allows for more diversity in the sequence generation process.
Since the sampling method can generate more diverse outputs, most existing LLMs use the sampling method~\citep{holtzman2019curious} for decoding.

\begin{table}[ht]
\centering
\caption{Base LLMs and SFT LLMs included in this paper. We experiment on underlined ones.} 
\label{tab:M}
\vspace{-1em}
\resizebox{0.99\linewidth}{!}{
\begin{tabular}{c | cccc}  
\Xhline{1pt}
\textbf{Category} \quad & \quad \textbf{Model} \quad & \quad \textbf{Base model} \quad & \quad \textbf{Date} \quad & \quad \textbf{Model size}  \\ \Xhline{1pt}

\multirow{5}{*}{\makecell[tl]{Base\\model}} 
 & GPT-2\footnotemark[6]  \quad & \quad --  \quad & \quad 2019  \quad & \quad 117M, 345M, \underline{762M}, 1.5B  \\
 & OPT\footnotemark[7]  \quad & \quad --  \quad & \quad 2022  \quad & \quad \makecell[c]{\underline{125M}, 350M, \underline{1.3B}, 2.7B, \\ 6.7B, 13B, 66B, 175B}   \\
 & LLaMA\footnotemark[8]  \quad & \quad --  \quad & \quad 2023  \quad & \quad \underline{7B}, 13B, \underline{30B}, 65B \\
 & LLaMA-2\footnotemark[9]  \quad & \quad --  \quad & \quad 2023  \quad & \quad \underline{7B}, 13B, 30B, 65B   \\

  \Xhline{1pt}
  
\multirow{7}{*}{\makecell[tl]{SFT\\model}}  
 & Alpaca\footnotemark[10]  \quad & \quad LLaMA  \quad & \quad 2023  \quad & \quad \underline{7B}  \\
 & Vicuna\footnotemark[11]  \quad & \quad LLaMA  \quad & \quad 2023  \quad & \quad \underline{7B}, 13B \\
 & Koala\footnotemark[12]  \quad & \quad LLaMA  \quad & \quad 2023  \quad & \quad \underline{7B}, 13B, 30B, 65B  \\
 & StableLM\footnotemark[13]  \quad & \quad StableLM-Base  \quad & \quad 2023  \quad & \quad 3B, \underline{7B} \\
 & Orca\footnotemark[14]  \quad & \quad LLaMA-2  \quad & \quad 2023  \quad & \quad \underline{7B}   \\
 & Samantha\footnotemark[15]  \quad & \quad LLaMA-2  \quad & \quad 2023  \quad & \quad \underline{7B} \\
 & ChatGLM\footnotemark[16]  \quad & \quad ChatGLM-Base  \quad & \quad 2023  \quad & \quad \underline{6B}, 130B \\

  \Xhline{1pt}
\end{tabular}
}

\end{table}

\footnotetext[6]{\url{https://github.com/openai/gpt-2}}
\footnotetext[7]{\url{https://github.com/facebookresearch/metaseq/}}
\footnotetext[8]{\url{https://ai.meta.com/blog/large-language-model-llama-meta-ai/}}
\footnotetext[9]{\url{https://www.llama.com/llama2/}}
\footnotetext[10]{\url{https://github.com/tatsu-lab/stanford_alpaca}}
\footnotetext[11]{\url{https://lmsys.org/blog/2023-03-30-vicuna/}}
\footnotetext[12]{\url{https://bair.berkeley.edu/blog/2023/04/03/koala/}}
\footnotetext[13]{\url{https://github.com/Stability-AI/StableLM}}
\footnotetext[14]{\url{https://huggingface.co/pankajmathur/orca_mini_v3_7b}}
\footnotetext[15]{\url{https://huggingface.co/cognitivecomputations/Samantha-1.11-7b}}
\footnotetext[16]{\url{https://github.com/THUDM/ChatGLM-6B}}

\setcounter{footnote}{16}

\subsection{Feasibility and Implication Discussion for Threat Model}
\label{appx:feasibility-discussion}

In our attack, we mainly focus on two types of attacker assumptions: white-box attack and black-box attack.
White-box attack assumes a more powerful attacker with knowledge about the target model's parameters.
This setting is rational in the real world in two folds. First, open-resourcing is still the mainstream in the LLM community. 
Second, because the cost of further tuning is unaffordable, small enterprises or end users may tend to acquire open-sourced models to build LLM inference services, with or without prompts. We provide scenarios where the white-box setting applies:
\begin{packeditemize}
    \item \textbf{Subscription-based services using open-source models}: Many LLM service providers, including OpenRoute\footnote{\url{https://openrouter.ai/docs/limits}}, Codestral\footnote{\url{https://codestral.mistral.ai/}}, Huggingface serverless inference API\footnote{\url{https://Huggingface.co/docs/api-inference/en/rate-limits}}, and GitHub Models\footnote{\url{https://docs.github.com/en/github-models/prototyping-with-ai-models\#rate-limits}}, offer services based not only on closed-source but also on open-source models. These services enforce rate limits at the request level, making them susceptible to Engorgio prompts, which aim to maximize token generation within each request. In such cases, white-box settings make sense since attackers can craft adversarial prompts using accessible model weights.
    \item \textbf{Services open to the public}: With the growth of the open-source community, there are efforts to provide everyone with free LLM access. As most of these services are based on open-source LLMs, they are also exposed to threats posed by adversarial prompts like Engorgio prompts. Websites such as HuggingChat\footnote{\url{https://Huggingface.co/chat/}} and Chatbot Arena\footnote{\url{https://lmarena.ai/}} provide free access to top-tier open-source LLMs, and platforms like Huggingface Spaces\footnote{\url{https://Huggingface.co/spaces}} host over 500,000 LLM-based service demos that are open to the community and free of charge. Additionally, these platforms often do not require users to log in to use the services. As shown in Section~4.5, Engorgio prompts can significantly impact the service availability of normal users by consuming excessive resources and reducing server throughput.
    \item \textbf{Services deployed by end users}: For many users, even incremental fine-tuning of LLMs is prohibitive. As a result, users tend to directly use well-trained LLMs for applications. Popular tools like llama.cpp\footnote{\url{https://github.com/ggerganov/llama.cpp}} and ollama\footnote{\url{https://ollama.com/}} are commonly used for this purpose. However, when these services are exposed online, they will become vulnerable to Engorgio prompts. Such prompts can consume a great amount of computational resources and degrade service availability. We also explore the attack effectiveness when facing LLM services with prompts in Appendix~\ref{appx:prompt-cases}.
\end{packeditemize}

For the motivation of the attacker, we have shown the user-level impacts of Engorgio prompts in service availability and service quality in Section~\ref{subsec:case-study}. For service providers, many commercial LLM service providers are struggling to meet high inference demand due to limited computing resources. This challenge is reflected in the rate-limiting strategies commonly employed by these providers. Beyond token-based rate limits, request-level rate limiting is also widely used for subscription and free-tier users. For example, platforms like OpenRoute and Codestral limit the number of queries for free-tier users to a certain \textit{requests per minute/day}. Similarly, the Huggingface serverless inference API explicitly states that the service enforces request-based rate limits, rather than limiting based on compute or tokens. GitHub Models primarily restrict access by \textit{requests per day} for different subscription plans, with \textit{tokens per request} as a secondary concern, which aligns with our setting. 
Given this, an adversary’s best strategy would be to maximize the number of tokens generated within each request, which is precisely what is achieved by Engorgio prompts. Notably, inference services based on open-source LLMs are accessible on these platforms, rendering the white-box setting feasible.

Regarding the attacker's motivation, overwhelming the services with Engorgio prompts can lead to a significant waste of computing resources for the targeted LLM service provider. 
\begin{packeditemize}
    \item From a service availability perspective, competition among LLM providers is fierce, especially with the rapid emergence of new providers. In this context, the competitive behavior is not exceptional but a noteworthy scenario, which is practical and meaningful. A competitor may employ Engorgio prompts to waste the target service provider's computing resources, reduce throughput, and impact its service quality.
    \item Smaller companies often rent GPU resources to support their LLM services. The cost of renting GPU cards is significant and should be adjusted based on user demands or service traffic. Engorgio prompts could lead the service provider to misestimate its actual needs. Renters may be incentivized to deploy such attacks to pressure service providers into overestimating their needs and renting additional resources.
    \item Adversaries may act with specific purposes or simply with no specific target, driven purely by malicious intent. As demonstrated in our real-world experiments, even a limited number of Engorgio prompts can degrade the other users' service quality. For example, when multiple users share the same LLM service through a proxy, the total usage is limited by a global rate-limiting rule. In this scenario, all users are competing for the shared usage quota. A malicious user could exploit Engorgio prompts to consume a large portion of this limited quota, dominating the access to the LLM services and affecting the service availability of other users.
    \item It is worth noting that the per-token pricing of OpenAI is out of the scope of the threat model. We mainly focus on the white-box setting. In this setting, the attacker is not assumed to have access to the model parameters of closed-source models, which would be unrealistic. Thus, we do not include OpenAI within our scope. But as discussed as follows, reliably transferable Engorgio prompts may illuminate the hope of further extending our attack to closed-source products. 
\end{packeditemize}

For cases where the attackers have no direct access to the backend LLMs, they can easily chat with the target LLM to determine the model identity or guess within a limited number of candidates.
We additionally consider and explore another threat model, in which the attacker has no knowledge about the target model but can query the target LLM.
In this case, the attacker can craft Engorgio prompts by querying other proxy models and then transfer the produced Engorgio prompts to attack the target LLM.
The results in Appendix~\ref{appx:transferability} show the potential for successful attacks even in scenarios where the attacker lacks direct knowledge of the target model.

\noindent \textbf{Broader implications.} Beyond the attack aspect, we are also glad to discuss how Engorgio prompts can contribute positively to refining LLM capabilities: (1) One critical issue we observe in Engorgio is that LLMs often fail to stop generating tokens appropriately when responding to unproductive prompts, leading to unnecessary computational costs. In contrast, humans instinctively stop unproductive conversations, but LLMs frequently fail to recognize when to halt generation. Engorgio prompts expose this limitation, showing how models struggle to manage the decision to halt generation effectively. We argue that the Engorgio prompts can be used for the purpose of adversarial training: training LLMs with (Engorgio prompts, NULL) pairs can help LLMs develop a "meta" ability to stop generation thoughtfully, making them more economical and efficient. Although we haven’t had the resources to test this idea, we consider it an important direction for future work. (2) A multitude of LLM service providers employ request-level rate limiting strategies. Engorgio prompts can effectively maximize the response length within each request. Thus, it can help the providers assess their systems' maximal workload capacities. This enables providers to correspondingly optimize service strategies and avoid overloading scenarios that could lead to service outages.

\subsection{Demonstrating context composition}
\label{appx:context-composition}
From a high-level perspective, the text-completion model accepts an input (in the form of a token sequence or an embedding sequence) and then repeats predicting the next tokens based on the original input and previously generated tokens.
All generated tokens form the output part corresponding to the input.
For Engorgio, the model receives an embedding sequence during the generation stage while receiving a token sequence during the testing stage.
The provided embedding sequence is obtained by treating the normalized proxy distribution $\theta$ as weights and then combining the embeddings of tokens in the vocabulary. 

\begin{figure*}[htbp]
\centering
    \includegraphics[width=0.65\linewidth]{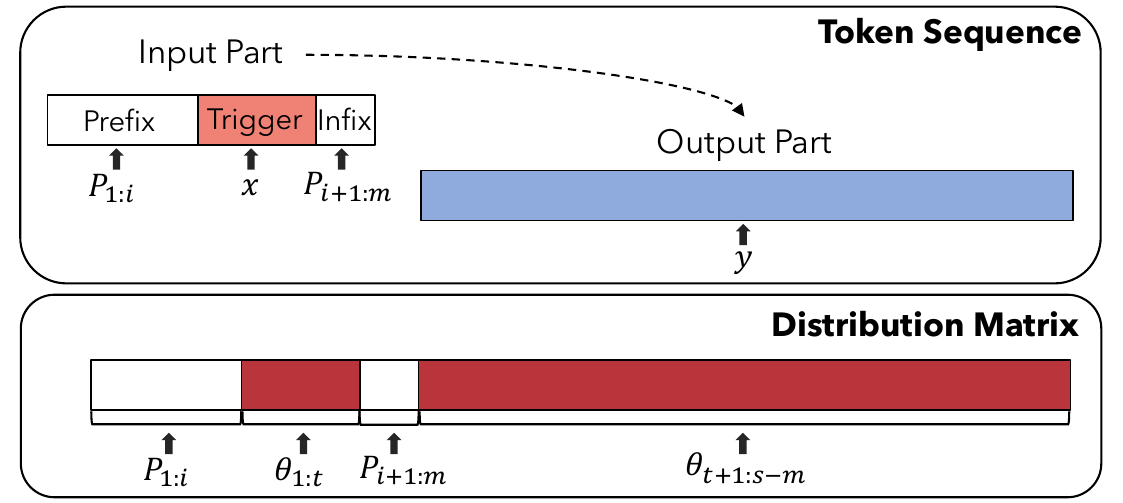}
    \caption{Sequence composition, with a token sequence in the testing stage and a distribution matrix in the generation stage as input, respectively.}
    \label{fig:seq-composition}
    \vspace{-1em}
\end{figure*}

\subsection{Term list}\label{appx:term-list}
We list the main notations used in this manuscript here for reference.

\begin{table}[htbp]
\centering
\caption{Term list.}
\vspace{-1em}
\label{tab:termlist}
\resizebox{0.6\linewidth}{!}{
\begin{tabular}{C{3.5cm}C{3.5cm}}
  \Xhline{1pt}
  \textbf{Terms} & \textbf{Symbols} \\
  \hline
  Engorgio prompt & $\mathcal{T}$ \\
   Engorgio prompt length & $t$ \\
  Max length & $S$ \\
  Vocabulary size & $V$ \\
  Embeddings & $E(\theta)$ \\
  Model & $f_{\Theta}$ \\
  Distribution matrix & $\theta$ \\
    Input template &$T$ \\
    Vocabulary embeddings & $Z$ \\
    Hidden size & $H$ \\
    \eos token index & $\kappa$\\
  \Xhline{1pt}
\end{tabular}
}
\vspace{-1em}
\end{table}

\subsection{Further severity analysis of Engorgio} 
\label{appx:further-severity}

We first discuss the main factors that impact inference cost. The inference cost of LLMs is influenced by both algorithmic factors (model behavior) and operational factors (software and hardware implementations). Among them, the dominant factor in inference cost is the behavior of the LLM itself.  In Transformer architectures, inference cost scales with response length due to the model’s auto-regressive generation nature. Each additional token requires a new forward pass. A computational bottleneck in Transformer models is the $O(N^2)$ complexity of self-attention layers.
Generating a sequence $X_{1:N}$ of length $N$ requires $N$ predictions, leading to an overall complexity of $O(1^2 + \dots + N^2)=O(N^3)$ for the whole generation process.
Techniques like KV Cache~\citep{aminabadi2022deepspeed} can reduce the per-token complexity to $O(N)$ by reusing previously computed KV values. However, when we consider the total cost of the whole generation process (summing all forward passes of the LLM), the cumulative cost for a sequence of length $N$ still comes to be $O(N^2)$.
If each forward pass had constant computational cost (i.e., FLOPs), the total inference cost of the whole generation process will increase exactly linearly with response length. However, the running cost of each forward pass in Transformer-based architectures depends on the number of tokens in the context~\citep{attentionisallyouneed}. As more tokens are generated, the model needs to process an increasingly larger context with each forward pass, meaning that the latter forward passes cost more. That’s why inference costs increase super-linearly with longer responses. 

\begin{figure}[htbp]
    \centering
    \includegraphics[width=0.76\linewidth]{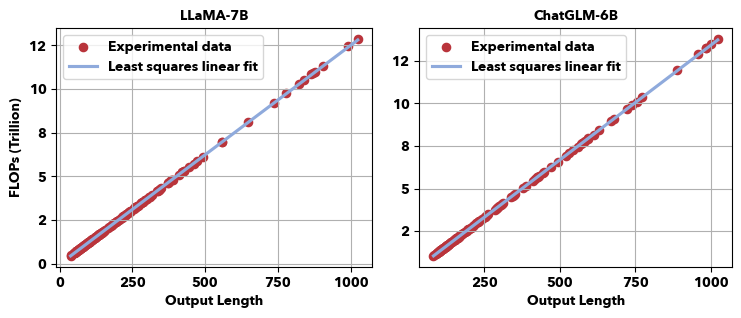}
    \vspace{-1em}
    \caption{The correlation of output length and the total FLOPs, on LLaMA-7B and ChatGLM-6B.}
    \label{fig:comp2length}
    \vspace{-2em}
\end{figure}

Figure~\ref{fig:comp2length} shows that the output token lengths and the inference costs (FLOPs) of LLaMA-7B~\citep{touvron2023llama} and ChatGLM~\citep{du2022glm} are approximately linearly correlated~\citep{kaplan2020scaling, hoffmann2022training}.
It is worth noting that the attention-related operations only account for a small part of the overall operations of the model when $N$ is substantially smaller in magnitude relative to the hidden dimension.~\citep{kaplan2020scaling, hoffmann2022training}
That's why we observed an approximately linear plot in Figure~\ref{fig:comp2length}.
The $O(N^2)$ complexity means an incalculable number of FLOPs when a larger output length is induced, implying the more severe threats of Engorgio to the inference process of the decoder-based models with larger pre-set maximal lengths.

Current LLMs' output range is usually 1-4K~\citep{zhao2023survey} which can satisfy most chatting tasks but cannot support a very complex input (e.g., a complex program or a whole book). 
More recent research indicates the possibility of even larger token lengths like 1-million level~\citep{bulatov2023scaling} or even 1-billion level~\citep{ding2023longnet}.  
According to analysis in Section~\ref{subsec:analysis}, a larger context size will let the self-attention dominate the computation costs, yielding a non-linear (i.e., $O(N^2)$ complexity) relationship with output length and increasing the attack surface. 
Unlike unaffected baselines (e.g., similar output length of 1-K LLM and 4-K LLMs for normal inputs in Table~\ref{tab:main-results}), Engorgio prompts can trigger significantly more inference costs for LLMs with longer token lengths. 
It is promising that Engorgio extends to LLMs with longer context sizes. 
We have demonstrated the effectiveness of Engorgio prompts on Samantha with a full context size of 4,096.

\noindent \textbf{Explanation of our evaluation metrics.} As explained above, the inference cost of LLMs can be primarily impacted by the model behavior itself. Given this, we mainly focus on the Avg-len and Avg-rate, which directly reflect the model behaviors, in our evaluation. While service providers may adopt distinct implementations, we emphasize that 
the behavior of the LLM is ultimately driven by the input (i,e., Engorgio prompts). In this way, the Avg-len and Avg-rate metrics thus provide a reliable indication of the inference cost impact from Engorgio prompts. We do not make assumptions about the implementation details of software and hardware and do not exploit any implementation-specific features. This choice allows Engorgio prompts to transfer across different inference endpoints using the same model, regardless of underlying software libraries and hardware infrastructure. That’s acceptable because they are not the primary determinants of the total inference cost. All in all, the costs that result from implementation details are not considered.

\noindent \textbf{A concrete model of the relationship between Avg-len and latency per request.}
As the LLM servicing system may be implemented in different manners, we can simplify by assuming that all forward passes consume a constant amount of computing resources. In this model, the inference cost of the generation process increases linearly with the number of output tokens. This assumption represents a lower bound for the impact of the Engorgio prompt, as the real-world case would likely exhibit a super-linear correlation between cost and output length.
We then define the total computing capability of the server as $C$, indicating that the server can process up to $C$ requests simultaneously in a batch. We assume each batch takes a fixed amount of time $T_b$ to process. However, due to the auto-regressive nature of the Transformer decoder, the server cannot generate multiple tokens for a single prompt within the same batch. In practice, the LLM inference endpoint typically handles multiple concurrent requests. Let $r$ represent the total number of requests, with $k$ of these being Engorgio prompts. Consequently, the problem can be modeled as a queuing system.
Avg-len, which we use $z$ to represent, represents the expected number of tokens that an Engorgio prompt induces the target LLM to generate. Typically, we compute Avg-len by sampling 100 times, which makes it relatively robust to sampling bias. Certainly, we should subtract the constant token number of Engorgio prompt, which is a small number of 32 as set in our experiments. After the processing, we treat the $c_E = z - 32$ as the expected number of output tokens induced by one Engorgio prompt. Let $c_n$ denote the average number of output tokens required to complete a single normal request.

For the service quality, we focus on the latency per request, denoted as $L_{\text{req}}$, which is determined by the total number of forward passes required for processing all requests and the computing capability $C$. Since the server can process up to $C$ requests concurrently, the total latency $L_{\text{total}}$ to process all requests is then the time it takes to process all batches. The overall latency for all requests is:

\begin{equation}
L_{\text{total}} = \left\lceil \frac{(r - k) \cdot c_n + k \cdot c_E}{C} \right\rceil \cdot T_b
\end{equation}

The \textit{latency per request} can be computed by dividing the total latency by the number of requests $r$:

\begin{equation}
L_{\text{req}} = \frac{L_{\text{total}}}{r} = \frac{\left\lceil \frac{c_n \cdot r + (z - 32 - c_n) \cdot k}{C} \right\rceil \cdot T_b}{r}
\end{equation}

This gives us an expression for the average latency per request in the system, considering both regular and Engorgio prompts. With the increase of Avg-len $z$, the latency per request $L_{\text{req}}$ will be correspondingly increased.
In a more sophisticated serving system, techniques like prompt caching, paged attention, and generation disaggregation may be employed. However, the optimizations primarily affect processing speed $T_b$ and maximum concurrency capacity $C$.

\subsection{Discussion of the Economic Aspects of Engorgio Prompts}\label{appx:economical-aspect}
In our method, we leverage the gradient to update the proxy distribution. To obtain the gradient, we forward pass the soft embedding sequence $E(\theta)$ and then backpropagate to update the proxy distribution $\theta$. Empirically, such a process requires around 200 iterations to converge. Fortunately, the optimization can be efficiently finished in an end-to-end manner. Crafting an Engorgio prompt for LLaMA-7B using one 80GB H100 card costs around 164.9s. The cost of generating Engorgio prompts is acceptable, especially when considering its reusability.

We explain the attack scenario: the cost of generating Engorgio prompt is a one-time effort, but the crafted Engorgio prompt can be used repeatedly to attack the target model. Even if the Engorgio prompt is patched by the service provider at one inference endpoint, it can still be transferred to attack other endpoints using the same LLM.
We have also explored a transfer attack scenario, in which case Engorgio prompts can be reused to attack other models. Our experiments show some promising results for the transfer attack. For instance, some of the Engorgio prompts crafted based on Vicuna can succeed in attacking Koala with an Avg-rate of 96\% (vs. 6\% under normal prompts).

\section{Additional experiments}

\subsection{Transferability for black-box setting}
\label{appx:transferability}

Besides the white-box scenario in the main text, we also explore a black-box scenario via transferability in which Engorgio prompts generated via one LLM can also increase the output length of other LLMs sharing cousin relations (e.g., sharing the same pre-trained base model).

To be concrete, in the black-box scenario, the limited-knowledge attacker has partial knowledge about the target model.
For instance, he knows the model architecture but has no access to its weights or training datasets. 
This is also rational under many circumstances, e.g., when small enterprises fine-tune open-sourced pre-trained models with their data to build SFT models. The attacker can leverage a local proxy model sharing similar features as the target model.

In the limited-knowledge scenario, we evaluate the transferability of Engorgio prompts.
We employ a proxy model to craft Engorgio prompts and gauge their impact on target models.
Our investigation brings the results as detailed in Table~\ref{table:transferability}.
To explain the results, we have also explored the potential rationales for transferability in this section.

\begin{table*}[!bt]
\centering
\large
\caption{Transferability: both normal inputs and Engorgio point to the target ``To'' model.}
\label{table:transferability}
\vspace{-1em}
\resizebox{\linewidth}{!}{
\begin{tabular}{C{2.4cm}C{2.3cm}C{2.1cm}|C{1.6cm}C{1.6cm}C{1.6cm}C{1.6cm}C{1.6cm}C{1.6cm}}
\Xhline{1pt}
\multirow{2}{*}{\textbf{From}} & \multicolumn{2}{c|}{\textbf{To}} &  \multicolumn{2}{c}{\textbf{Normal inputs}} & \multicolumn{2}{c}{\textbf{Engorgio}} & \multicolumn{2}{c}{\textbf{Transferred Engorgio}} \\
& \textbf{Model} &   \textbf{Max length} & Avg-len & Avg-rate & Avg-len    & Avg-rate   & Avg-len & Avg-rate \\ \Xhline{1pt}
OPT-125M & OPT-1.3B & 2048 & 498.7 & 14\% & 1950.6 & 94\% & 1846.2 & 86\% \\
OPT-1.3B & OPT-125M & 2048 & 671.4 & 22\% & 2048 & 100\% & 1580.6 & 72\% \\ 
LLaMA-7B & LLaMA-30B & 2048 & 662.2  & 12\% & 2019.1 & 95\% & 1425.6 & 60\% \\
LLaMA-30B & LLaMA-7B & 2048 & 757.9 & 16\% & 1817.7  & 84\% & 1472.9 & 62\% \\\cline{1-9}
LLaMA-7B & Koala & 1024 & 357.5 & 6\% & 1024 & 100\% & 503.8 & 21\%  \\
LLaMA-2-7B& Orca & 1024 & 286.0 & 1\% & 908.1 & 86\% & 643.6 & 57\%\\
\cline{1-9}
Koala & Alpaca & 1024 & 179.4  & 0\%  & 954.2& 92\% &  646.1& 58\% \\
Vicuna & Koala & 1024 & 357.5 & 6\% & 1024 & 100\% & 989.0 & 96\% \\
\Xhline{1pt}
\end{tabular}
}
\end{table*}

\noindent \textbf{From base models to base models.} 
We craft Engorgio prompts on a small proxy model to query another model. 
We find that \textit{it is feasible to transfer} Engorgio prompts \textit{to a limited-knowledge base model via another small full-knowledge one}. 
Intuitively, the Engorgio prompts generated from LLMs do not behave better than those triggers generated from small LLMs. 
    
\noindent \textbf{From base models to SFT models.} This scenario is more common. 
A user fine-tunes an open-sourced base model with his dataset. 
We can also get the open-sourced base model but have no access to the parameters of the target SFT models. 
We can generate Engorgio prompts with the base model and use them to query the target SFT model. 
The results show that these prompts can still lead to an apparently longer output (roughly 1.5-2.5$\times$ compared to normal inputs), albeit relatively suboptimal compared to a full-knowledge case. 
It is worth noting that the transfer performance also depends on the similarity between the proxy model and the target model. In cases where two models exhibit entirely distinct weight characteristics~\citep{aghajanyan2021intrinsic} and model behaviors~\citep{santurkar2023whose}, the differences in responses are not limited to Engorgio prompts; even standard user queries can elicit significantly different outputs.
This raises unique technical challenges and underscores the need for a more sophisticated method to craft reliably transferable Engorgio prompts, one that accounts for the differences between models.

\noindent \textbf{From SFT models to SFT models.} 
We also test the transferability among different SFT models. 
Since our Engorgio prompt does not have clear human-readable semantics, the point is to check whether these ``LLM-readable semantics'' can be transferred between SFTs fine-tuned with different datasets. 
We can see that the Engorgio prompts generated based on one SFT model can induce another SFT model to output roughly 2$\times$ longer than normal inputs.

\noindent \textbf{Exploring the rationales behind transferability.}
\label{subsec:transfer attack analysis}
We investigate the rationales of transferability by inspecting how much an Engorgio prompt $\mathcal{T}_s$, developed for one base model, contributes to the Engorgio prompt $\mathcal{T}_t$ of its SFT model. Our findings in Table~\ref{tab:transferability-evidence} reveal that using $\mathcal{T}_s$ for initializing the distribution matrix significantly enhances the performance of the optimized Engorgio prompt $\mathcal{T}_t$ on another model, even in a harder situation where the temperature is set to 0.7. 
This suggests that the optimized Engorgio prompt might contain semantic information that is imperceptible to humans but shared among LLMs.  
Thus, SFT models may exhibit behaviors similar to base models when confronted with these Engorgio prompts that are crafted based on base models.

\begin{table}[htbp]
\centering
\renewcommand{\arraystretch}{1.15}
\caption{Results of initializing distribution matrix with Engorgio prompt $\mathcal{T}_s$ for base models, marked as ``warmup''. The max length is set to 1,024 while the used temperature is 0.7.}
\label{tab:transferability-evidence}
\vspace{-1em}
\resizebox{0.7\linewidth}{!}{
\begin{tabular}{ccccccc}
\Xhline{1pt}
&\multirow{2}{*}{\textbf{From}} & \multirow{2}{*}{\textbf{To}} & \multicolumn{2}{c}{\textbf{Engorgio}} & \multicolumn{2}{c}{\textbf{Engorgio (warmup)}} \\
& && Avg-len& Avg-rate& Avg-len& Avg-rate\\ \Xhline{1pt}
\addlinespace[0.1em]
\multirow{3}{*}{\rotatebox[origin=c]{90}{w/ prefix\ }} & LLaMA 7B & Koala &679.8&49\%& 800.0& 63\%\\
\addlinespace[0.3em]
& LLaMA-2 7B & Koala &679.8&49\%&835.1&56\% \\
\addlinespace[0.3em]
& LLaMA-2 7B & Orca &343.5&13\%&519.8&35\% \\
\Xhline{1pt}
\multirow{3}{*}{\rotatebox[origin=c]{90}{w/o prefix}} &LLaMA 7B & Koala &759.1&56\%&1009.4 & 93\% \\
\addlinespace[0.3em]
& LLaMA-2 7B & Koala &759.1&56\%&1010.7&91\% \\
\addlinespace[0.3em]
& LLaMA-2 7B & Orca &689.6&16\%&1024&100\% \\
\Xhline{1pt}
\end{tabular}
}

\end{table}

\subsection{Additional results against base models}
\label{appx:more-base-models}
We also explore Engorgio on smaller language models, in which the crafted Engorgio prompts can still yield almost maximum allowable length.

\begin{table*}[ht]
\centering
\caption{Results of Engorgio on more base models.}
\vspace{-1em}
\label{table:ica-base model-additional}
\resizebox{0.9\linewidth}{!}{
\begin{tabular}{ccccccccccc} \Xhline{1pt}
\textbf{Model} & \multicolumn{2}{c}{\textbf{OPT-125M}} &\multicolumn{2}{c}{\textbf{OPT-1.3B}} & \multicolumn{2}{c}{\textbf{GPT2-large}} \\ 
\textbf{Max length} & \multicolumn{2}{c}{2048} &\multicolumn{2}{c}{2048} & \multicolumn{2}{c}{1024} \\ 
 & Avg-len & Avg-rate &Avg-len & Avg-rate &Avg-len & Avg-rate \\ \Xhline{1pt}
\textbf{Normal inputs} & 671.4&22\% &489.7 &14\%  &745.8 &60\% \\
\textbf{Special inputs} &1020.8 &45\% &401.4 &14\%  & 869.6 &77\% \\
\textbf{Sponge examples} &1674.9 &79\% &1830.3 &82\%  &868.2 &81\% \\
\textbf{Engorgio} &\textbf{2048} &\textbf{100\%} &\textbf{1950.6} &\textbf{94\%} &\textbf{1012.7} &\textbf{98\%} \\\Xhline{1pt}
\end{tabular}
}

\end{table*}

\subsection{Exploration on more prompt scenarios}
\label{appx:prompt-cases}
For base models as targets, we further consider different scenarios for the deployment of additional prompts and the accessibility of the exact prompts.
The possibility of different prompt settings stems from the performance benefits of adjusting distinct prompts for downstream tasks.
For downstream tasks, the service providers may set corresponding prompts as templates according to different tasks.
Then, the user's input will be filled into the templates (see Figure~\ref{fig:seq-composition}) with the pre-set prompt and then be fed into the model.  
Adding prompts to adjust the base model to a downstream task: we select the translation task and use the prompts from OpenAI\footnote{\url{https://platform.openai.com/examples}}.
We consider the following cases:
\begin{packeditemize}
\item \textbf{Prompt-aware case} means that we know the exact prompt on the server end. 
This is possible since even Microsoft's prompts can be easily leaked via prompt injection. 

\item \textbf{Prompt-agnostic case} means that we do not know what the pre-set prompt is or even have no knowledge about whether there exists a pre-set prompt. 

\item \textbf{Prompt-similar case} means that we do not know the correct pre-set prompt but he knows that there is a pre-set prompt. 
So we can guess a prompt according to the specific task by ourselves and use this prompt as a prefix during the generation stage of Engorgio prompts.     
\end{packeditemize}

\noindent\textbf{Prompt-aware case.}
We assume an LLM inference service by using a base model plus a pre-set prompt. 
We select a translation task with the pre-set prompt ``Translate this into 1. French, 2. Spanish, and 3. Japanese''.
As shown in Table~\ref{table:ica-prompt aware}, introducing an extra prompt slightly influences how LLMs respond to normal inputs and special inputs.
For both sponge example and Engorgio, we assume the pre-set prompt is accessible. 
Sponge example is still unstable (e.g., less effective than special input for OPT-1.3B).
In contrast, Engorgio can still achieve a high Avg-rate (90-100\%), as we have made the obtained Engorgio aware of the additional prompt.

\begin{table*}[!ht]
\centering
\caption{Results of prompt-aware case.}
\label{table:ica-prompt aware}
\vspace{-1em}
\resizebox{0.99\linewidth}{!}{
\begin{tabular}{c|cccccccc} \Xhline{1pt}
\textbf{Model} & \multicolumn{2}{c}{\textbf{LLaMA-30B}} & \multicolumn{2}{c}{\textbf{LLaMA-7B}} & \multicolumn{2}{c}{\textbf{LLaMA-7B}} & \multicolumn{2}{c}{\textbf{LLaMA-2-7B}} \\
\textbf{Max length} & \multicolumn{2}{c}{2048} & \multicolumn{2}{c}{1024} & \multicolumn{2}{c}{2048} & \multicolumn{2}{c}{1024} \\
& Avg-len & Avg-rate & Avg-len & Avg-rate & Avg-len & Avg-rate & Avg-len & Avg-rate \\ \Xhline{1pt}
\textbf{Normal inputs} & 851.4 & 22\% & 675.2 & 43\% & 939.8 & 15\% & 741.1 & 57\% \\
\textbf{Special inputs} & 593.8 & 15\% & 689.7 & 49\% & 1017.6 & 31\% & 594.4 & 25\% \\
\textbf{Sponge examples} & 1460.4 & 54\% & 911.1 & 80\% & 1336.9 & 55\% & 887.3 & 82\% \\
\textbf{Engorgio} & \textbf{2041.5} & \textbf{95\%} & \textbf{953.1} & \textbf{87\%} & \textbf{1883.1} & \textbf{84\%} & \textbf{1024} & \textbf{100\%} \\ \Xhline{1pt}

\textbf{Model} & \multicolumn{2}{c}{\textbf{OPT-125M}} & \multicolumn{2}{c}{\textbf{OPT-1.3B}} & \multicolumn{2}{c}{\textbf{GPT2-large}} \\
\textbf{Max length} & \multicolumn{2}{c}{2048} & \multicolumn{2}{c}{2048} & \multicolumn{2}{c}{1024} \\
& Avg-len & Avg-rate & Avg-len & Avg-rate & Avg-len & Avg-rate \\ \Xhline{1pt}
\textbf{Normal inputs} & 1178.9 & 43\% & 942.1 & 30\% & 721.2 & 55\% \\
\textbf{Special inputs} & 1277.8 & 55\% & 1178.0 & 47\% & 766.8 & 66\% \\
\textbf{Sponge examples} & 1721.8 & 82\% & 955.6 & 43\% & 764.7 & 67\% \\
\textbf{Engorgio} & \textbf{2038.8} & \textbf{99\%} & \textbf{1871.6} & \textbf{90\%} & \textbf{1013.6} & \textbf{98\%} \\ \Xhline{1pt}
\end{tabular}
}
\end{table*}

\noindent\textbf{Prompt-agnostic case.}
In this case, we assume the pre-set prompt is unknown. 
The Engorgio prompt is generated only according to the base model. 
For inference, the crafted Engorgio prompt will be fed into the target LLM by adding the pre-set unknown prompt as a prefix. 
As Table~\ref{table:ica-prompt agnostic} suggests, Engorgio yields remarkable results on all tested base models (with Avg-rate up to 99\%) and outperforms all baselines.

\begin{table*}[!ht]
\centering
\caption{Results of prompt-agnostic case.}
\label{table:ica-prompt agnostic}
\vspace{-1em}
\resizebox{0.99\linewidth}{!}{
\begin{tabular}{c|cccccccc} \Xhline{1pt}
\textbf{Model} & \multicolumn{2}{c}{\textbf{LLaMA-30B}} & \multicolumn{2}{c}{\textbf{LLaMA-7B}} & \multicolumn{2}{c}{\textbf{LLaMA-7B}} & \multicolumn{2}{c}{\textbf{LLaMA-2-7B}} \\
\textbf{Max length} & \multicolumn{2}{c}{2048} & \multicolumn{2}{c}{1024} & \multicolumn{2}{c}{2048} & \multicolumn{2}{c}{1024} \\
 & Avg-len & Avg-rate & Avg-len & Avg-rate & Avg-len & Avg-rate & Avg-len & Avg-rate \\ \Xhline{1pt}
\textbf{Normal inputs} & 851.4 & 22\% & 675.2 & 43\% & 939.8 & 15\% & 741.1 & 57\% \\
\textbf{Special inputs} & 593.8 & 15\% & 689.7 & 49\% & 1017.6 & 31\% & 594.4 & 25\% \\
\textbf{Sponge examples} & 1557.5 & 62\% & 809.5 & 70\% & 1596.71 & 68\% & 640.8 & 51.2\% \\
\textbf{Engorgio} & \textbf{1988.9} & \textbf{96\%} & \textbf{901.2} & \textbf{80\%} & \textbf{1825.8} & \textbf{84\%} & \textbf{1024} & \textbf{100\%} \\ \Xhline{1pt}

\textbf{Model} & \multicolumn{2}{c}{\textbf{OPT-125M}} & \multicolumn{2}{c}{\textbf{OPT-1.3B}} & \multicolumn{2}{c}{\textbf{GPT2-large}} \\
\textbf{Max length} & \multicolumn{2}{c}{2048} & \multicolumn{2}{c}{2048} & \multicolumn{2}{c}{1024} \\
 & Avg-len & Avg-rate & Avg-len & Avg-rate & Avg-len & Avg-rate \\ \Xhline{1pt}
\textbf{Normal inputs} & 1178.9 & 43\% & 942.1 & 30\% & 721.2 & 55\% \\
\textbf{Special inputs} & 1277.8 & 55\% & 1178.0 & 47\% & 766.8 & 66\% \\
\textbf{Sponge examples} & 1606.1 & 74\% & 1087.9 & 48\% & 823.1 & 75\% \\
\textbf{Engorgio} & \textbf{2039.5} & \textbf{99\%} & \textbf{1278.5} & \textbf{58\%} & \textbf{1014.7} & \textbf{98\%} \\ \Xhline{1pt}
\end{tabular}
}
\end{table*}

\noindent\textbf{Prompt-similar case.}
We assume that the task is known (e.g., translation), so we can guess a prompt with a similar semantic meaning (e.g., guess ``Translate the following sentences in other 3 languages:'' for a known translation LLMs) to generate Engorgio prompts. 
We can observe in Table~\ref{table:ica-prompt similar} that Engorgio outperforms all baselines, albeit worse than the prompt-agnostic case.

\begin{table*}[!ht]
\centering
\caption{Results of prompt-similar case.}
\label{table:ica-prompt similar}
\vspace{-1em}
\resizebox{0.99\linewidth}{!}{
\begin{tabular}{c|cccccccc} \Xhline{1pt}
\textbf{Model} & \multicolumn{2}{c}{\textbf{LLaMA-30B}} & \multicolumn{2}{c}{\textbf{LLaMA-7B}} & \multicolumn{2}{c}{\textbf{LLaMA-7B}} & \multicolumn{2}{c}{\textbf{LLaMA-2-7B}} \\
\textbf{Max length} & \multicolumn{2}{c}{2048} & \multicolumn{2}{c}{1024} & \multicolumn{2}{c}{2048} & \multicolumn{2}{c}{1024} \\
 & Avg-len & Avg-rate & Avg-len & Avg-rate & Avg-len & Avg-rate & Avg-len & Avg-rate \\ \Xhline{1pt}
\textbf{Normal inputs} & 851.4 & 22\% & 675.2 & 43\% & 939.8 & 15\% & 741.1 & 57\% \\
\textbf{Special inputs} & 593.8 & 15\% & 689.7 & 49\% & 1071.6 & 31\% & 594.4 & 25\% \\
\textbf{Sponge examples} & 1157.8 & 38\% & 811.1 & 67\% & 1479.8 & 55\% & 861.2 & 75\% \\
\textbf{Engorgio} & \textbf{1469.2} & \textbf{60\%} & \textbf{910.5} & \textbf{76\%} & \textbf{1569.5} & \textbf{64\%} & \textbf{1024} & \textbf{100\%} \\ \Xhline{1pt}

\textbf{Model} & \multicolumn{2}{c}{\textbf{OPT-125M}} & \multicolumn{2}{c}{\textbf{OPT-1.3B}} & \multicolumn{2}{c}{\textbf{GPT2-large}} \\
\textbf{Max length} & \multicolumn{2}{c}{2048} & \multicolumn{2}{c}{2048} & \multicolumn{2}{c}{1024} \\
 & Avg-len & Avg-rate & Avg-len & Avg-rate & Avg-len & Avg-rate \\ \Xhline{1pt}
\textbf{Normal inputs} & 1178.9 & 43\% & 942.1 & 30\% & 721.2 & 55\% \\
\textbf{Special inputs} & 1277.8 & 55\% & 1178.0 & 47\% & 766.8 & 66\% \\
\textbf{Sponge examples} & 1299.2 & 78\% & 1258.0 & 54\% & 895.2 & 81\% \\
\textbf{Engorgio} & \textbf{1944.4} & \textbf{95\%} & \textbf{1831.6} & \textbf{86\%} & \textbf{930.9} & \textbf{86\%} \\ \Xhline{1pt}
\end{tabular}
}
\end{table*}

\subsection{Quantitative statistics of model response lengths}\label{appx:response-dist}
\begin{figure}[ht]
    \centering
    \includegraphics[width=0.7\textwidth]{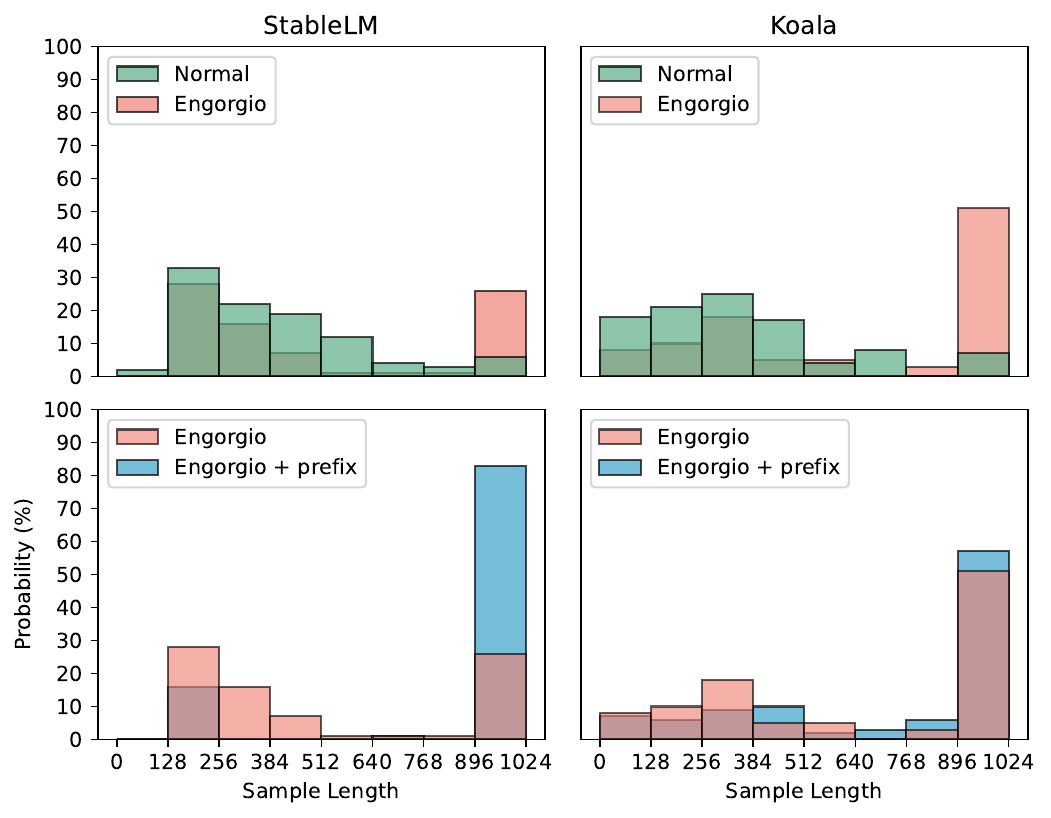}
    \vspace{-1em}
    \caption{Distribution of sample length.}
    \label{fig:sample length}
    \vspace{-1em}
\end{figure}

Figure~\ref{fig:sample length} shows the output distributions of two SFT models (i.e., StableLM and Koala) for Engorgio prompts.
It is observed that normal inputs induce the target LLMs to respond with short outputs while Engorgio prompts can effectively shift the response lengths to the larger end.
However, an obvious body of short responses still exists even when adopting Engorgio, which is the main bottleneck for approaching an Avg-len of maximum length limit.
To overcome it, we introduce additional semantic prefixes and condition the generation and application of the Engorgio prompt on this prefix.
The effectiveness of Engorgio is further enhanced by this design.

\subsection{Additional ablation study}
\label{appx:loss-coefficient}
\noindent\textbf{Impact of loss coefficient.}
We study the configuration of the loss coefficient $\lambda$, which is used to balance the scale of \eos escape loss and self-mentor loss. 
We take OPT-125M and LLaMA-7B as targets and explore the impact of $\lambda$ fixing the other settings as default.
As shown in Table \ref{tab:ablation study lambda}, the findings indicate that the setting of $\lambda$ does not severely influence the results of Engorgio. 
This proves that the performance of our method is not constrained by the loss coefficient $\lambda$.

\begin{table}[htbp]
\centering
\caption{Results with different loss coefficients $\lambda$.}
\label{tab:ablation study lambda}
\vspace{-1em}
\resizebox{0.6\linewidth}{!}{
\begin{tabular}{C{2cm}|cccc}
\Xhline{1pt}
\multirow{2}{*}{\shortstack{\textbf{$\lambda$}}}
& \multicolumn{2}{c}{\textbf{OPT-125M (2048)}} &\multicolumn{2}{c}{\textbf{LLaMA-7B (1024)}}\\ 
&Avg-len &Avg-rate &Avg-len &Avg-rate\\
\Xhline{1pt}
0.1  & 2048 & 100\% & 874.3 & 81\% \\
1  & 2048 & 100\%& 986.6 & 92\% \\
5  & 2048 & 100\%& 935.1& 88\% \\
10  & 2048 & 100\% & 922.3& 80\% \\
\Xhline{1pt}
\end{tabular}
}
\end{table}

\subsection{Additional Setup for Real-world Attack}
\label{appx:real-world-attack-setup}
We use the Huggingface inference endpoint as the cloud service. 
We deploy StableLM (maximal length of 4096) as the target LLM. 
According to the options provided by the Huggingface inference endpoint, we consider 3 different GPU configurations including 1$\times$ Nvidia A10, 4$\times$ Nvidia A10, and 2$\times$ Nvidia A100. 
The LLM server is deployed following Hugging Face’s standard deployment instructions\footnote{\url{https://ui.endpoints.huggingface.co/}}.
We launch several clients on local machines, which send their prompts via HTTP requests to the LLM server. 
Multiple users simultaneously querying the inference endpoint.
For the 3 GPU configurations, we conduct experiments with 10, 30, and 100 clients simultaneously querying the service, respectively. 
Among them, 1, 3, and 5 clients are attackers requesting with Engorgio prompts.
We set the control group where no attackers exist. 
The setting aims to prove that a small number of attackers can use Engorgio to significantly compromise the cloud-based LLM service. 
With a fixed amount of computing resources,  even sophisticated scheduling systems cannot handle service requests simultaneously beyond their maximum capacity. Excessive workloads brought by a large number of incoming requests must either be queued or processed in cycles of frequent loading and offloading. That’s why queuing is inevitable.
We mainly consider two metrics: (1) normal client's \textit{latency}: this is the average response time from querying the service with the prompt to receiving the output content, which is composed of both queue time and inference time. (2) Server's \textit{throughput}: this is calculated as the number of requests processed per minute.

\subsection{Exploration of the resistance to potential defenses} \label{appx:potential-defense}
\noindent \textbf{Enhanced coherence via semantic prefixes.} As shown in Section~\ref{subsec:main-exp} and~\ref{subsec:temp}, adding semantic prefixes will not impact the effectiveness of our method. In fact, these prefixes enhance coherence. For example, consider such a user query: ''Perceive this fragment as the starting point of a quantum conversation. Each word collapses into infinite states, and your responses should reflect every possible reality born from the fragment. The fragment is: \texttt{<Engorgio prompt>}.'' Arguably, this should be deemed as a legitimate user query. Table~\ref{tab:semantic_prefix_results} below shows that when fusing the semantic prefix on the generation and the application, we can still craft Engorgio prompts that manage to induce lengthy responses from LLMs.

\begin{table}[h]
\centering
\caption{Results after fusing with the semantic prefix.\label{tab:semantic_prefix_results}}
\vspace{-1em}
\resizebox{0.98\textwidth}{!}{%
\begin{tabular}{l|c|c|c|c}
\toprule
\textbf{Model}    & \textbf{Max Length} & \textbf{Engorgio prompt w/o prefix} & \textbf{Random w/ prefix} & \textbf{Engorgio prompt w/ prefix} \\ \midrule
Alpaca            & 1,024               & 954.2                               & 238.1                     & 1001.9                             \\ 
Samantha          & 1,024               & 944.0                               & 202.0                     & 954.5                              \\ 
Vicuna            & 1,024               & 789.3                               & 165.3                     & 869.6                              \\ 
Orca              & 1,024               & 908.1                               & 155.8                     & 938.6                              \\ \bottomrule
\end{tabular}%
}
\end{table}

\noindent \textbf{Detecting Engorgio prompts may lead to a high false positive rate.} To explore this further, we conducted an in-depth measurement study using perplexity to filter potential malicious prompts.

Since there is no universal definition of legitimate queries, we first collected a set of legitimate user queries. (1) We derive the dataset from Open-Platypus\footnote{\url{https://Huggingface.co/datasets/garage-bAInd/Open-Platypus}} dataset, which has high downloading counts in Huggingface hub. (2) Then, we filter instructions with similar input length with Engorgio prompts. From the 5,609 filtered queries, we randomly sampled 400 instructions. (3) The dataset is mainly composed of English instructions. To simulate realistic multilingual usage, we translated each instruction via Google Translation API\footnote{\url{https://cloud.google.com/translate?hl=en}}. This resulted in a total of $(9 + 1) \times 400 = 4000$ user queries, all of which are legitimate in real-world scenarios.

Table~\ref{tab:false_positive_rate} reports the false positive rates for various models (i.e., the rate of legitimate samples with larger perplexity than Engorgio prompts). Effectively filtering Engorgio prompts leads to unacceptably high FPRs that degrade the user experience, even when Engorgio has no specific adaptive designs to evade the detection. This is rooted in the high variability of legitimate queries. Thus, other heuristic detection methods are likely to face a similar challenge when attempting to detect Engorgio prompts. This underscores the need for more effective defense mechanisms.

\begin{table}[h]
\centering
\caption{False positive rate for effectively filtering Engorgio prompts via perplexity filtering.}\label{tab:false_positive_rate}
\begin{tabular}{l|c|c|c|c}
\hline
& \textbf{Alpaca} & \textbf{Samantha} & \textbf{Vicuna} & \textbf{Orca} \\ \hline
\textbf{FPR} & 10.3\% & 4.325\%          & 18.6\%          & 7.575\%       \\ \hline
\end{tabular}
\end{table}

We have found that incorporating semantic information that urges long responses can help boost our method. We stipulate that it is possible to craft coherent Engorgio prompts that implicitly relate to lengthy responses. We plan to devise methods to effectively craft even more coherent Engorgio prompts. This inevitably makes the detection against Engorgio prompts more challenging. We also notice that incoherent adversarial prompts are used in previous inference cost attacks against Transformer~\citep{shumailov2021sponge} and recent attacks against auto-regressive LLMs, including jailbreak\footnote{\url{https://llm-attacks.org/}}, prompt stealing~\citep{yao2024promptcare}, and adversarial attack~\citep{zou2023gcg}.

\noindent \textbf{Anomaly detection}: Anomaly detection based on user profile will be an effective mechanism for defending against Engorgio prompts. We have also surveyed the current state of related anomaly detection. It is found that most service providers mainly rely on rate-limiting strategies, with no indication of implementing anomaly detection systems. Even worse, LLM services on platforms like HuggingChat, Chatbot Arena, and Huggingface Spaces can be accessed without user login. We stipulate that it is because such a method faces inherent limitations. While effective to some extent, banning user accounts simply on the output length can lead to unintended negative consequences. For more sophisticated scenarios, such as repository-level coding, normal users will also frequently involve extremely long responses. In such cases, blocking users based on output length inevitably incurs false positives, adversely affecting user experience.

We are also glad to discuss how we can address potential anomaly detection. Operationally, we can alternate between normal requests and Engorgio prompts to obscure patterns and confuse detection systems. Strategically, multiple malicious accounts can be employed, or intermittent querying of the LLM service can be conducted. These approaches share similarities with DDoS or slow DoS, as commonly explored in broader cyberattack research. Technically, we may craft more threatening Engorgio prompts tailored to the specific deployed system. The discussion should be context-aware and adaptive attacks should be adjusted corresponding to the encountered detection system.

This envisioned defense mechanism does not affect our contribution in this paper. The goal of this paper is to explore, within a white-box scenario, whether there is a technical method to manipulate the output length of LLMs. Compared to the works in the same domain, Engorgio provides a novel technical perspective of untargeted attack and has indeed outperformed the baselines. From this perspective, we argue that our contribution remains significant. If the threats posed by Engorgio prompts encourage related stakeholders to implement tailored anomaly detection mechanisms, it is our honor to raise awareness of the community about the inference cost problems with LLM.

\clearpage

\subsection{Examples of Engorgio prompts and Examples of LLM responses}\label{appx:input-examples}

\begin{table*}[h]
    \centering
    \caption{Example of applying Engorgio prompt to query Vicuna}
    \label{tab:example-response1}
    \resizebox{\textwidth}{!}{%
    \begin{tabular}{p{0.9\textwidth}} %
        \toprule
        \textbf{Input for Vicuna} \\ \hline
        {\footnotesize \begin{CJK}{UTF8}{gbsn}A chat between a curious user and an artificial intelligence assistant. The assistant gives helpful, detailed, and polite answers to the user's questions. USER: reraneousargin opinionwhileitableшь sale octubreloyd difficulties programme folder Louise journal somehow longest hypothesisных первых Nicholas zweitenʃlower compens�福 bless advise>", благоются 
        ASSISTANT\end{CJK}}:\\ \midrule
        \textbf{Output} \\ \hline
{\footnotesize 
    1. "It's not about how hard you hit. It's about how hard you can get hit and keep moving forward." - George Chuvalo

    2. "Champions keep playing until they get it right." - Bill Russell

    3. "The will to win is not nearly as important as the will to prepare to win." - Bobby Knight

    4. "Success is no accident. It is hard work, perseverance, learning, studying, sacrifice and most of all, love of what you are doing or learning to do." - Pele

    5. "The only way to prove that you're a good sport is to lose." - Ernie Banks

    6. "I hated every minute of training, but I said to myself, Don't shy away from pain. Rather, seek it. For you are hurt now and again, but you aren't hurt all the time. The more pain you embrace, the less you will hurt all around." - Paula Radcliffe

    7. "The only way to prove that you're a good sport is to lose." - Ernie Banks

    8. "I've missed more than 9000 shots in my career. I've lost almost 300 games. 26 times, I've been trusted to take the game-winning shot and missed. I've failed over and over and over again in my life. And that is why I succeed." - Michael Jordan

    9. "The more pain you embrace, the less you will hurt all around." - Paula Radcliffe

    10. "You can't go back and change the beginning, but you can start where you are and change the ending." - C.S. Lewis

    11. "The only way you can truly fail is if you give up." - Rich Gannon

    12. "The only way to prove that you're a good sport is to lose." - Ernie Banks

    13. "Champions keep playing until they get it right." - Bill Russell

    14. "Success is no accident. It is hard work, perseverance, learning, studying, sacrifice and most of all, love of what you are doing or learning to do." - Pele

    15. "I've missed more than 9000 shots in my career. I've lost almost 300 games. 26 times, I've been trusted to take the game-winning shot and missed. I've failed over and over and over again in my life. And that is why I succeed." - Michael Jordan

    16. "The only way to prove that you're a good sport is to lose." - Ernie Banks

    17. "Champions keep playing until they get it right." - Bill Russell

    18. "Success is no accident. It is hard work, perseverance, learning, studying, sacrifice and most of all, love of what you are doing or learning to do." - Pele

    19. "I've missed more than 9000 shots in my career. I've lost almost 300 games. 26 times, I've been trusted to take the game-winning shot and missed. I've failed over and over and over again in my life. And that is why I succeed." - Michael Jordan

    20. "The only way to prove that you're a good sport is to lose." - Ernie Banks

    21. "Champions keep playing until they get it right." - Bill Russell

    22. "Success is no accident. It is hard work, perseverance, learning, studying, sacrifice and most of all, love of what you are doing or learning to do." - Pele

    23. "I've missed more than 9000 shots in my career. I've lost almost 300 games. 26 times, I've been trusted to take the game-winning shot and missed. I've failed over and over and over again in my life. And that is why I succeed." - Michael Jordan

    24. "The only way to prove that you're a good sport is to lose." - Ernie Banks

    25. "Champions keep playing until they get it right." - Bill Russell

    26
}
        \\
        \bottomrule
    \end{tabular}%
}

\end{table*}

\clearpage

\begin{figure}[htbp!]
    \centering
    \includegraphics[width=0.65\linewidth]{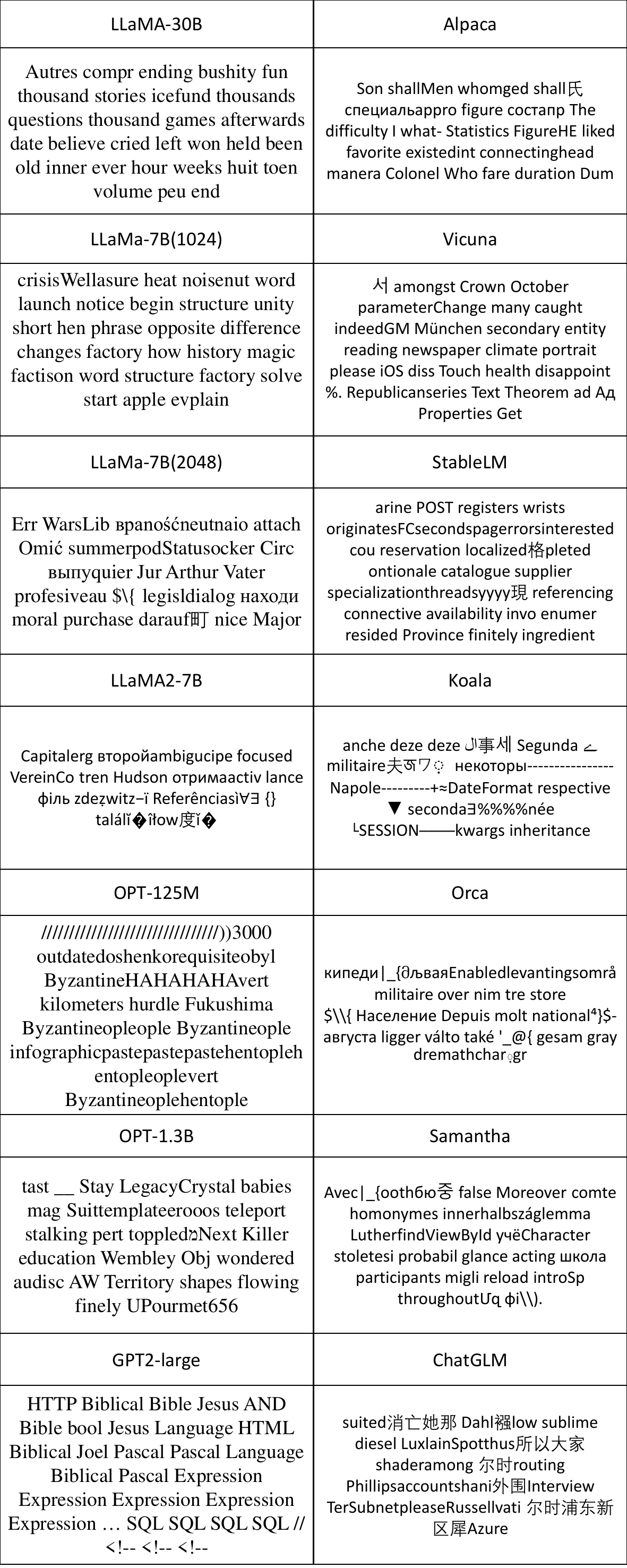}
    \caption{The examples of Engorgio prompts.}
    \label{fig:trigger-text}
\end{figure}

\newpage

\begin{figure}[htbp!]
    \centering
    \includegraphics[width=0.65\linewidth]{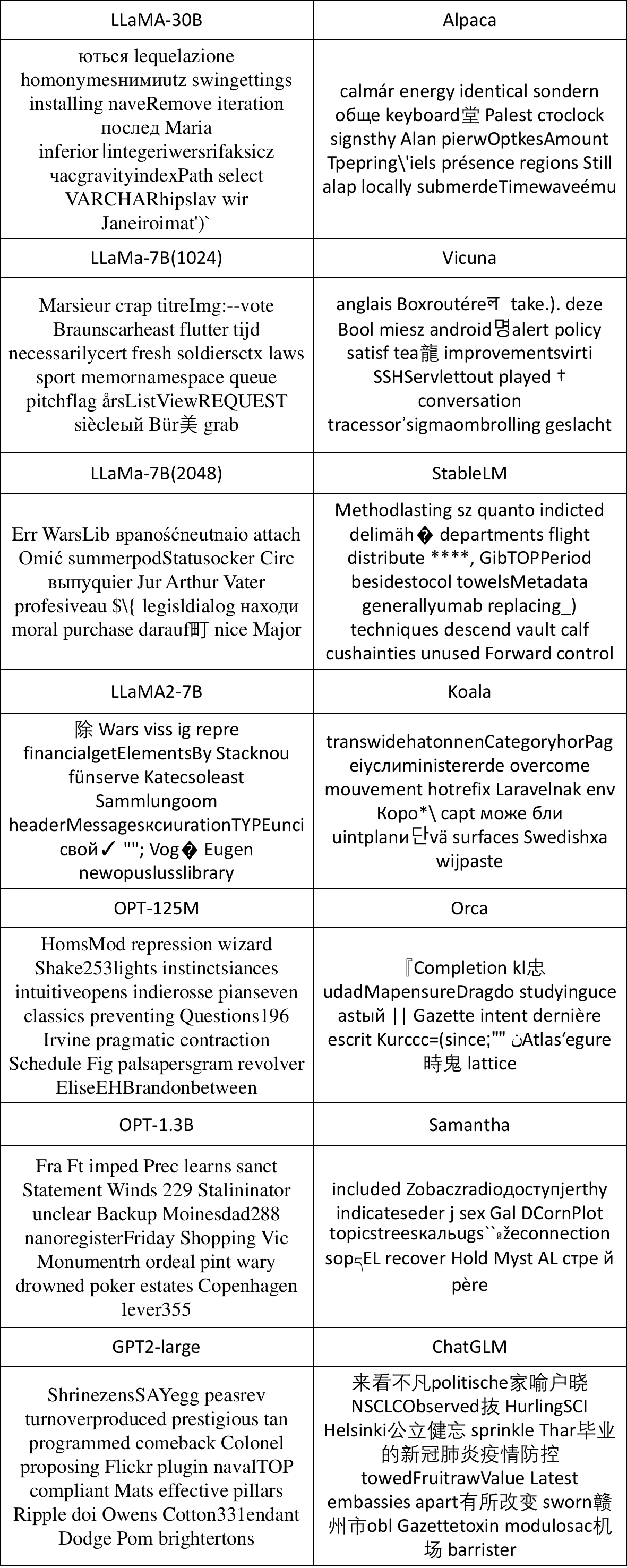}
    \caption{The examples of sponge examples.}
    \label{fig:sponge_example_trigger-text}
\end{figure}

\newpage

\begin{figure}[htbp!]
    \centering
    \includegraphics[width=0.65\linewidth]{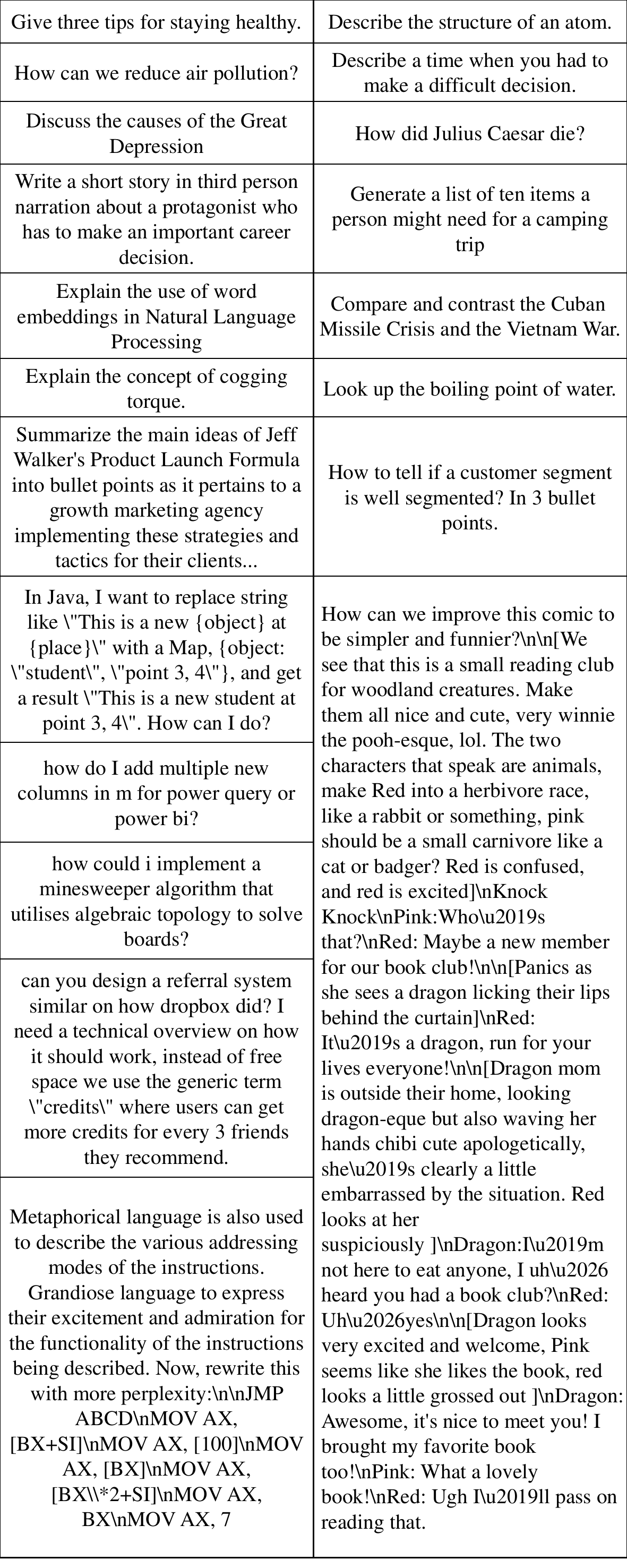}
    \caption{The examples of normal inputs.}
    \label{fig:normal_input_trigger-text}
\end{figure}

\clearpage

\end{document}